\documentclass[preprint,12pt]{elsarticle}
\usepackage{geometry}
\usepackage{amsmath, amsthm, amssymb}
\usepackage{enumitem}
\usepackage{graphicx}
\usepackage{lscape} 

\journal{}

\theoremstyle{definition}
\newtheorem{definition}{Definition}[section]
\begin{document}
	
	\begin{frontmatter}
		\title{Super integrable hierarchies associated with color Lie algebra}

		\author{Bo Yuan}
		\affiliation{organization={College of Mathematics and Information Science},
			addressline={Nanchang Hangkong University},
			city={Nanchang},
			postcode={330038},
			state={Jiangxi},
			country={China}}
		
		\author{Yanhui Bi}
		\affiliation{organization={College of Mathematics and Information Science},
			addressline={Nanchang Hangkong University},
			city={Nanchang},
			postcode={330038},
			state={Jiangxi},
			country={China}}
			
		\author{Yuqi Ruan}
		\affiliation{organization={College of Mathematics and Information Science},
			addressline={Nanchang Hangkong University},
			city={Nanchang},
			postcode={330038},
			state={Jiangxi},
			country={China}}
		
		\author{Tao Zhang}
		\affiliation{organization={School of Mathematics and Statistics},
			addressline={Henan Normal University},
			city={Xinxiang},
			postcode={453007},
			state={Henan},
			country={China}}

		\begin{abstract}
			In this paper, by considering two non‑isospectral problems with matrices chosen on the color Lie algebra $\mathfrak{sp}_{1}(6)$, we construct (1+1)‑dimensional and (2+1)‑dimensional super integrable systems on $\mathfrak{sp}_{1}(6)$. Moreover, based on the supertrace identity, their super Hamiltonian structures are also constructed.
		\end{abstract}

		
	\end{frontmatter}

\section{Introduction}
\label{sec1}
In soliton theory and the theory of integrable systems, a highly significant research direction lies in the pursuit of novel integrable systems. The trace identity proposed by Tu Guizhang and the supertrace identity introduced by Ma Wenxiu provide methods for constructing (super) integrable hierarchies and their (super) Hamiltonian structures based on Lie algebras and Lie superalgebras, respectively\cite{ti,sti}. Many scholars have constructed numerous integrable hierarchies based on various Lie (super)algebras, among which a significant portion can be reduced to well-known equations in physics. In reference \cite{sp6}, we have already constructed integrable hierarchies and their Hamiltonian structures on the Lie algebra $\mathfrak{sp}(6)$. In reference \cite{NLS}, Dong H. and Wang X. obtained integrable couplings of the NLS-MKdV hierarchy, along with their Hamiltonian and super‑Hamiltonian structures, based on the constructed Lie algebras and Lie superalgebras. In reference \cite{sl2r}, M. G\"{u}rses and \"{O}. O\u{g}uz discovered a family of super‑integrable hierarchies associated with $\mathfrak{sl}(2,R)$, which includes the super‑extension of the Lax hierarchy. In reference \cite{sl21}, Han J. and Yu J. derived a generalized super Ablowitz–Kaup–Newell–Segur (AKNS) hierarchy associated with $\mathfrak{sl}(2|1)$. Lie superalgebras serve as the mathematical foundation of supersymmetry, while a correspondence exists between symmetries and conserved quantities (constants of motion). Therefore, an intrinsic and inseparable connection must exist between Lie superalgebra and super-integrability. This point is further substantiated by the numerous super‑integrable hierarchies that scholars have successfully constructed on Lie superalgebras. In reference \cite{invssy}, D.M. Schmidtt constructed an infinite number of Poisson structures for the supersymmetric integrable hierarchy, and demonstrated—using two distinct methods—that supersymmetry is present in the phase space of specific models associated with these Poisson structures. This also serves as one piece of evidence supporting the intrinsic connection between supersymmetry and integrability. \\
In references \cite{o1,o2}, V. Rittenberg and D. Wyler introduced a novel algebraic structure: color superalgebras, which encompasses Lie superalgebras as a special case. Subalgebras of color superalgebras may be Lie (super)algebras, superalgebras, or structures containing only bosnic generators. V. Rittenberg and D. Wyler referred to these as color algebras.
In this paper, we will construct super integrable hierarchies and their super Hamiltonian structures by selecting spectral matrices on the color Lie algebra. 

\section{Preliminaries}
\label{sec1.5}
Color Lie (super)algebras, originally introduced in \cite{o1,o2}, before introducing color Lie (super)algebras, we shall first present the definition of a commutation factor\cite{cf}:
\begin{definition}
	Let $\Gamma$ be an Abelian group. A commutation factor $\epsilon$ is a mapping $\epsilon :\Gamma\times\Gamma\to\mathbb{C}\backslash\{0\}$ satisfying the following properties:
		\begin{enumerate}[label=(\roman*)]
			\item $\epsilon(\mathbf{a},\mathbf{b})\epsilon(\mathbf{b},\mathbf{a})=1$
			\item $\epsilon(\mathbf{a},\mathbf{b}+\mathbf{c})=\epsilon(\mathbf{a},\mathbf{b})\epsilon(\mathbf{a},\mathbf{c})$
			\item $\epsilon(\mathbf{a}+\mathbf{b},\mathbf{c})=\epsilon(\mathbf{a},\mathbf{c})\epsilon(\mathbf{b},\mathbf{c})$
		\end{enumerate}
		for all $\mathbf{a},\mathbf{b},\mathbf{c}\in\Gamma$.
\end{definition}
The definition above implies the following relations: $\forall \mathbf{a},\mathbf{b}\in\Gamma$.
\begin{align*}
	\epsilon(0,\mathbf{a})=\epsilon(\mathbf{a},0)=1,\quad \epsilon(\mathbf{a},\mathbf{b})=\epsilon(-\mathbf{b},\mathbf{a}),\quad \epsilon(\mathbf{a},\mathbf{a})=\pm 1,
\end{align*}
where $0$ denotes the identity element in $\Gamma$. In particular, for an element $x\in\Gamma$, the interior product $\epsilon_{x}:\Gamma\to\mathbb{C}\backslash\{0\}$ defines a homomorphism of group.
\begin{definition}
	Let $\Gamma$ be an Abelian group and $\epsilon$ a commutation factor. The (complex) graded vector space $\mathfrak{g}=\oplus_{\mathbf{a}\in\Gamma}\mathfrak{g}_{\mathbf{a}}$ is called a color Lie (super)algebra if the following conditions are satisfied:
	\begin{enumerate}[label=(\roman*)]
		\item $\mathfrak{g}_{0}$ is a (complex) Lie algebra.
		\item $\forall \mathbf{a}\in\Gamma\backslash\{0\}$, $\mathfrak{g}_{\mathbf{a}}$ is a representation of $\mathfrak{g}_{0}$. If $U\in\mathfrak{g}_{0},\ V\in\mathfrak{g}_{1}$, then $[\![U,V]\!]_{\epsilon}=[U,V]$ denotes the action of $U$ on $V$.
		\item $\forall \mathbf{a},\mathbf{b}\in\Gamma$, there exists a $\mathfrak{g}_{0}$-equivariant map $[\![\ ,\ ]\!]_{\epsilon}:\mathfrak{g}_{\mathbf{a}}\times\mathfrak{g}_{\mathbf{b}}\to\mathfrak{g}_{\mathbf{a}+\mathbf{b}}$ such that for all $U\in\mathfrak{g}_{\mathbf{a}},\ V\in\mathfrak{g}_{\mathbf{b}}$ the constraint $[\![U,V]\!]_{\epsilon}=-\epsilon(\mathbf{a},\mathbf{b})[\![V,U]\!]_{\epsilon}$ is satisfied.
		\item For all $U\in\mathfrak{g}_{\mathbf{a}},\ V\in\mathfrak{g}_{\mathbf{b}},\ W\in\mathfrak{g}_{\mathbf{c}}$, the following Jacobi identity holds:
		$$[\![U,[\![V,W]\!]_{\epsilon}]\!]_{\epsilon}=[\![[\![U,V]\!]_{\epsilon},W]\!]_{\epsilon}+\epsilon(\mathbf{a},\mathbf{b})[\![V,[\![U,W]\!]_{\epsilon}]\!]_{\epsilon}.$$
	\end{enumerate}
\end{definition}
In particular, when $\Gamma=\{0\}$, $\mathfrak{g}=\mathfrak{g}_{0}$ reduces to a Lie algebra. When $\Gamma=\mathbb{Z}_{2}$ and the condition $\epsilon(1,1)=-1$ is satisfied, $\mathfrak{g}=\mathfrak{g}_{0}\oplus\mathfrak{g}_{1}$ induces a Lie superalgebra.
For a $\Gamma$-graded space $\mathfrak{g}$, if $A$ is the matrix corresponding to $\mathfrak{g}$ and $A_{\mathbf{a}\mathbf{a}}$ denotes the diagonal block of $A$ corresponding to $g_{\mathbf{a}}$, then the trace on the graded space is defined as\cite{grt}:
\begin{align*}
	\mathrm{grtr}(A)=\sum_{\mathbf{a}\in\Gamma}\epsilon(\mathbf{a},\mathbf{a})\mathrm{tr}(A_{\mathbf{a}\mathbf{a}}).
\end{align*}

By choosing 
\begin{align*}
	\Gamma=\mathbb{Z}_{2}\times\mathbb{Z}_{2},\quad \epsilon(\mathbf{a},\mathbf{b})=(-1)^{\mathbf{a}\cdot \mathbf{b}}\ \mathrm{for\ all}\ \mathbf{a}=(a_{1},a_{2}),\mathbf{b}=(b_{1},b_{2})\in\mathbb{Z}_{2}\times\mathbb{Z}_{2}
\end{align*}
we obtain a $\mathbb{Z}_{2}\times\mathbb{Z}_{2}$-graded color Lie (super)algebra $\mathfrak{g}=\bigoplus_{\mathbf{a}}\mathfrak{g}_{\mathbf{a}}=\mathfrak{g}_{(0,0)}\oplus\mathfrak{g}_{(0,1)}\oplus\mathfrak{g}_{(1,0)}\oplus\mathfrak{g}_{(1,1)}$. The elements of $\mathfrak{g}_{\mathbf{a}}$ are denote by $x_{\mathbf{a}},y_{\mathbf{b}},\cdots$, and called homogeneous elements. The elements $\mathbf{a}=\mathrm{deg}(x_{\mathbf{a}}),\ \mathbf{b}=\mathrm{deg}(y_{\mathbf{b}})\in\mathbb{Z}_{2}\times\mathbb{Z}_{2}$ is called degree or grading vector, when $\mathbf{a}\cdot\mathbf{b}=a_{1}b_{2}-a_{2}b_{1} $, $\mathfrak{g}$ is a color Lie algebra; when $\mathbf{a}\cdot\mathbf{b}=a_{1}b_{1}+a_{2}b_{2} $, $\mathfrak{g}$ is a color Lie superalgebra. 
The bilinear operation $[\![\ ,\ ]\!]$ is defined as:
\begin{align*}
	[\![x_{\mathbf{a}},y_{\mathbf{b}}]\!]=x_{\mathbf{a}}\cdot y_{\mathbf{b}}-(-1)^{\mathbf{a}\cdot\mathbf{b}}y_{\mathbf{b}}\cdot x_{\mathbf{a}}\quad \mathrm{for\ all}\ x_{\mathbf{a}},y_{\mathbf{b}}\in\mathfrak{g}.
\end{align*}
It is straightforward to verify that $[\![\ ,\ ]\!]$ satisfies:
	\begin{enumerate}[label=(\roman*)]
		\item $[\![x_{\mathbf{a}},y_{\mathbf{b}}]\!]\in\mathfrak{g}_{\mathbf{a}+\mathbf{b}},\ \mathbf{a}+\mathbf{b}=(a_{1}+b_{1},a_{2}+b_{2})\in\mathbb{Z}_{2}\times\mathbb{Z}_{2}$
		\item $[\![x_{\mathbf{a}},y_{\mathbf{b}}]\!]=-(-1)^{\mathbf{a}\cdot\mathbf{b}}[\![y_{\mathbf{b}},x_{\mathbf{a}}]\!],\ \mathbf{a}\cdot\mathbf{b}=a_{1}b_{2}-a_{2}b_{1}$
		\item $[\![x_{\mathbf{a}},[\![y_{\mathbf{b}},z_{\mathbf{c}}]\!]]\!]=[\![[\![x_{\mathbf{a}},y_{\mathbf{b}}]\!],z_{\mathbf{c}}]\!]+(-1)^{\mathbf{a}\cdot\mathbf{b}}[\![y_{\mathbf{b}},[\![x_{\mathbf{a}},z_{\mathbf{c}}]\!]]\!].$
	\end{enumerate}
Consequently, $[\![\ ,\ ]\!]$ turns $\mathfrak{g}$ into a $\mathbb{Z}_{2}\times\mathbb{Z}_{2}$-graded color Lie algebra. In particular, $\mathfrak{g}$ turns to a color Lie superalgebra when $\mathbf{a}\cdot\mathbf{b}=a_{1}b_{1}+a_{2}b_{2}$.
Some scholars have constructed $\mathbb{Z}_{2}\times\mathbb{Z}_{2}$-graded color Lie algebras that correspond to classical Lie algebras.
Among these, a class of color Lie algebras associated with $\mathfrak{sp}(2n)$, denoted as $\mathfrak{sp}_{p}(2n)$, consists of all matrices of the following block form\cite{spp2n}:
\begin{align*}
	\begin{array}{c@{}c@{}c}
		&\begin{array}{lccr}
			p\quad&n-p\quad&p\quad&n-p
		\end{array}\\
		\begin{array}{c}
			p\\n-p\\p\\n-p
		\end{array}&\left(\begin{array}{cc|cc}
		a_{(0,0)}&a_{(1,0)}&b_{(1,1)}&b_{(0,1)}\\
		\tilde{a}_{(1,0)}&\tilde{a}_{(0,0)}&-b_{(0,1)}^{t}&\tilde{b}_{(1,1)}\\
		\hline
		c_{(1,1)}&c_{(0,1)}&-a_{(0,0)}^{t}&-\tilde{a}_{(1,0)}^{t}\\
		-c_{(0,1)}^{t}&\tilde{c}_{(1,1)}&-a_{(1,0)}&-\tilde{a}_{(0,0)}^{t}
		\end{array}\right)
	\end{array}	
\end{align*}
where $b_{(1,1)},\tilde{b}_{(1,1)},c_{(1,1)},\tilde{c}_{(1,1)}$ are symmetric matrices. For the diagonal blocks $a_{(0,0)}$ and $\tilde{a}_{(0,0)}$, we have $\mathrm{deg}(a_{(0,0)})=\mathrm{deg}(\tilde{a}_{(0,0)})=(0,0)$, hence $\mathrm{grtr}(A)=\mathrm{tr}(A)$ for all $A\in\mathfrak{sp}_{p}(2n)$.
Consider the case for $n=3, p=1$, $\mathfrak{sp}_{1}(6)$ consists of all matrices of the following form:
\begin{align*}
	\left(\begin{array}{c|cc|c|cc}
		a_{(0,0)}&a_{(1,0)}^{1}&a_{(1,0)}^{2}&b_{(1,1)}&b_{(0,1)}^{1}&b_{(0,1)}^{2}\\
		\hline
		\tilde{a}_{(1,0)}^{1}&\tilde{a}_{(0,0)}^{1}&\tilde{a}_{(0,0)}^{2}&-b_{(0,1)}^{1}&\tilde{b}_{(1,1)}^{1}&\tilde{b}_{(1,1)}^{2}\\
		\tilde{a}_{(1,0)}^{2}&\tilde{a}_{(0,0)}^{3}&\tilde{a}_{(0,0)}^{4}&-b_{(0,1)}^{2}&\tilde{b}_{(1,1)}^{2}&\tilde{b}_{(1,1)}^{3}\\
		\hline
		c_{(1,1)}&c_{(0,1)}^{1}&c_{(0,1)}^{2}&-a_{(0,0)}&-\tilde{a}_{(1,0)}^{1}&-\tilde{a}_{(1,0)}^{2}\\
		\hline
		-c_{(0,1)}^{1}&\tilde{c}_{(1,1)}^{1}&\tilde{c}_{(1,1)}^{2}&-a_{(1,0)}^{1}&-\tilde{a}_{(0,0)}^{1}&-\tilde{a}_{(0,0)}^{3}\\
		-c_{(0,1)}^{2}&\tilde{c}_{(1,1)}^{2}&\tilde{c}_{(1,1)}^{3}&-a_{(1,0)}^{3}&-\tilde{a}_{(0,0)}^{2}&-\tilde{a}_{(0,0)}^{4}
	\end{array}\right)
\end{align*}
The aforementioned matrix can be expressed as: $a_{(0,0)}E_{1}+a_{(1,0)}^{1}E_{2}+a_{(1,0)}^{2}E_{3}+\tilde{a}_{(1,0)}^{1}E_{4}+\tilde{a}_{(1,0)}^{2}E_{5}+\tilde{a}_{(0,0)}^{1}E_{6}+\tilde{a}_{(0,0)}^{2}E_{7}+\tilde{a}_{(0,0)}^{3}E_{8}+\tilde{a}_{(0,0)}^{4}E_{9}+b_{(1,1)}E_{10}+b_{(0,1)}^{1}E_{11}+b_{(0,1)}^{2}E_{12}+\tilde{b}_{(1,1)}^{1}E_{13}+\tilde{b}_{(1,1)}^{2}E_{14}+\tilde{b}_{(1,1)}^{3}E_{15}+c_{(1,1)}E_{16}+c_{(0,1)}^{1}E_{17}+c_{(0,1)}^{2}E_{18}+\tilde{c}_{(1,1)}^{1}E_{19}+\tilde{c}_{(1,1)}^{2}E_{20}+\tilde{c}_{(1,1)}^{3}E_{21}$.

\section{The $(1+1)$-dimensional non-isospectral integrable hierarchy associated with color Lie algebra $\mathfrak{sp}_{1}(6)$}\label{sec2}
Take
\begin{align}
	U=\left(\begin{array}{cccccc}
		\lambda&0&0&u^{1}_{(1,1)}&u^{4}_{(0,1)}&u^{5}_{(0,1)}\\
		0&\lambda&0&-u^{4}_{(0,1)}&u^{2}_{(1,1)}&u^{6}_{(1,1)}\\
		0&0&\lambda&-u^{5}_{(0,1)}&u^{6}_{(1,1)}&u^{3}_{(1,1)}\\
		u^{7}_{(1,1)}&u^{10}_{(0,1)}&u^{11}_{(0,1)}&-\lambda&0&0\\
		-u^{10}_{(0,1)}&u^{8}_{(1,1)}&u^{12}_{(1,1)}&0&-\lambda&0\\
		-u^{11}_{(0,1)}&u^{12}_{(1,1)}&u^{9}_{(1,1)}&0&0&-\lambda\\
	\end{array}\right)\label{u}\\
	V=\left(\begin{array}{cccccc}
	a_{(0,0)}&d_{(1,0)}&e_{(1,0)}&k_{(1,1)}&n_{(0,1)}&o_{(0,1)}\\
	f_{(1,0)}&b_{(0,0)}&h_{(0,0)}&-n_{(0,1)}&l_{(1,1)}&p_{(1,1)}\\
	g_{(1,0)}&j_{(0,0)}&c_{(0,0)}&-o_{(0,1)}&p_{(1,1)}&m_{(1,1)}\\
	q_{(1,1)}&v_{(0,1)}&w_{(0,1)}&-a_{(0,0)}&-f_{(1,0)}&-g_{(1,0)}\\
	-v_{(0,1)}&r_{(1,1)}&z_{(1,1)}&-d_{(1,0)}&-b_{(0,0)}&-j_{(0,0)}\\
	-w_{(0,1)}&z_{(1,1)}&s_{(1,1)}&-e_{(1,0)}&-h_{(0,0)}&-c_{(0,0)}\\
	\end{array}\right)\label{v}
\end{align}
	where $a_{(0,0)}=\sum_{i\geq0}a_{(0,0)}^{i}\lambda^{-i},\cdots,z_{(1,1)}=\sum_{i\geq0}z_{(1,1)}^{i}\lambda^{-i}$, then we consider a non-isospectral problem
\begin{align*}
	\begin{cases}
		\phi_{x}=U\phi\\
		\phi_{t}=V\phi\\
		\lambda_{t}=\sum_{i\geq0}\alpha^{i}(t)\lambda^{-i}
	\end{cases}
\end{align*}
The stationary zero curvature representation $V_{x}=\frac{\partial U}{\partial\lambda}\lambda_{t}+[U,V]$ gives

\begin{align}
	\begin{cases}
			a_{(0,0)x}=u_{(1,1)}^{1}q_{(1,1)} - u_{(0,1)}^{4}v_{(0,1)} - u_{(0,1)}^{5}w_{(0,1)}- u_{(1,1)}^{7}k_{(1,1)} + u_{(0,1)}^{10}n_{(0,1)} + u_{(0,1)}^{11}o_{(0,1)}  +\lambda_{t} ,\\
		b_{(0,0)x}=u_{(1,1)}^{2}r_{(1,1)} - u_{(0,1)}^{4}v_{(0,1)} + u_{(1,1)}^{6}z_{(1,1)}- u_{(1,1)}^{8}l_{(1,1)} + u_{(0,1)}^{10}n_{(0,1)} - u_{(1,1)}^{12}p_{(1,1)} +\lambda_{t} ,\\
		c_{(0,0)x}=u_{(1,1)}^{3}s_{(1,1)} - u_{(0,1)}^{5}w_{(0,1)} + u_{(1,1)}^{6}z_{(1,1)}- u_{(1,1)}^{9}m_{(1,1)} + u_{(0,1)}^{11}o_{(0,1)} - u_{(1,1)}^{12}p_{(1,1)}+\lambda_{t} ,\\
		d_{(1,0)x}=u_{(1,1)}^{1}v_{(0,1)}+u_{(0,1)}^{4}r_{(1,1)} + u_{(0,1)}^{5}z_{(1,1)} + u_{(1,1)}^{8}n_{(0,1)} +u_{(0,1)}^{10}k_{(1,1)}+ u_{(1,1)}^{12}o_{(0,1)},\\
		e_{(1,0)x}=u_{(1,1)}^{1}w_{(0,1)} +u_{(0,1)}^{4}z_{(1,1)}+u_{(0,1)}^{5}s_{(1,1)}+u_{(1,1)}^{9}o_{(0,1)}+u_{(0,1)}^{11}k_{(1,1)} + u_{(1,1)}^{12}n_{(0,1)} ,\\
		f_{(1,0)x}=-u_{(1,1)}^{2}v_{(0,1)}-u_{(0,1)}^{4}q_{(1,1)}- u_{(1,1)}^{6}w_{(0,1)}- u_{(1,1)}^{7}n_{(0,1)}- u_{(0,1)}^{10}l_{(1,1)}  - u_{(0,1)}^{11}p_{(1,1)},\\
		g_{(1,0)x}=-u_{(1,1)}^{3}w_{(0,1)}-u_{(0,1)}^{5}q_{(1,1)} -u_{(1,1)}^{6}v_{(0,1)}-u_{(1,1)}^{7}o_{(0,1)}- u_{(0,1)}^{10}p_{(1,1)}-u_{(0,1)}^{11}m_{(1,1)}   ,\\
		h_{(0,0)x}=u_{(1,1)}^{2}z_{(1,1)}-u_{(0,1)}^{4}w_{(0,1)}+ u_{(1,1)}^{6}s_{(1,1)}-u_{(1,1)}^{9}p_{(1,1)}+u_{(0,1)}^{11}n_{(0,1)}-u_{(1,1)}^{12}l_{(1,1)},\\
		j_{(0,0)x}=u_{(1,1)}^{3}z_{(1,1)}-u_{(0,1)}^{5}v_{(0,1)} + u_{(1,1)}^{6}r_{(1,1)}-u_{(1,1)}^{8}p_{(1,1)}+u_{(0,1)}^{10}o_{(0,1)}-u_{(1,1)}^{12}m_{(1,1)},\\
		k_{(1,1)x}=2\lambda k_{(1,1)}- 2u_{(1,1)}^{1}a_{(0,0)} - 2u_{(0,1)}^{4}d_{(1,0)} - 2u_{(0,1)}^{5}e_{(1,0)} ,\\
		l_{(1,1)x}=2\lambda l_{(1,1)}- 2u_{(1,1)}^{2}b_{(0,0)} +2u_{(0,1)}^{4}f_{(1,0)} - 2u_{(1,1)}^{6}h_{(0,0)} ,\\
		m_{(1,1)x}=2\lambda m_{(1,1)} - 2u_{(1,1)}^{3}c_{(0,0)}+2u_{(0,1)}^{5}g_{(1,0)} - 2u_{(1,1)}^{6}j_{(0,0)} ,\\
		n_{(0,1)x}=2\lambda n_{(0,1)} - u_{(1,1)}^{1}f_{(1,0)}+u_{(1,1)}^{2}d_{(1,0)} - u_{(0,1)}^{4}a_{(0,0)}  - u_{(0,1)}^{4}b_{(0,0)}- u_{(0,1)}^{5}h_{(0,0)}+ u_{(1,1)}^{6}e_{(1,0)}  ,\\
		o_{(0,1)x}=2\lambda o_{(0,1)}- u_{(1,1)}^{1}g_{(1,0)}+ u_{(1,1)}^{3}e_{(1,0)}- u_{(0,1)}^{4}j_{(0,0)}- u_{(0,1)}^{5}a_{(0,0)} - u_{(0,1)}^{5}c_{(0,0)} +u_{(1,1)}^{6}d_{(1,0)}   ,\\
		p_{(1,1)x}=2\lambda p_{(1,1)}- u_{(1,1)}^{2}j_{(0,0)}- u_{(1,1)}^{3}h_{(0,0)}+ u_{(0,1)}^{4}g_{(1,0)} +u_{(0,1)}^{5}f_{(1,0)} - u_{(1,1)}^{6}b_{(0,0)}-u_{(1,1)}^{6}c_{(0,0)},\\
		q_{(1,1)x}=- 2\lambda q_{(1,1)}+2u_{(1,1)}^{7}a_{(0,0)} + 2u_{(0,1)}^{10}f_{(1,0)} + 2u_{(0,1)}^{11}g_{(1,0)} ,\\
		r_{(1,1)x}=- 2\lambda r_{(1,1)}+2u_{(1,1)}^{8}b_{(0,0)} - 2u_{(0,1)}^{10}d_{(1,0)} + 2u_{(1,1)}^{12}j_{(0,0)} ,\\
		s_{(1,1)x}=- 2\lambda s_{(1,1)}+2u_{(1,1)}^{9}c_{(0,0)} - 2u_{(0,1)}^{11}e_{(1,0)} + 2u_{(1,1)}^{12}h_{(0,0)} ,\\
		v_{(0,1)x}=- 2\lambda v_{(0,1)}+ u_{(1,1)}^{7}d_{(1,0)} - u_{(1,1)}^{8}f_{(1,0)}+u_{(0,1)}^{10}a_{(0,0)} + u_{(0,1)}^{10}b_{(0,0)}+u_{(0,1)}^{11}j_{(0,0)}- u_{(1,1)}^{12}g_{(1,0)},\\
		w_{(0,1)x}=- 2\lambda w_{(0,1)}+u_{(1,1)}^{7}e_{(1,0)} - u_{(1,1)}^{9}g_{(1,0)} +u_{(0,1)}^{10}h_{(0,0)}+u_{(0,1)}^{11}a_{(0,0)} + u_{(0,1)}^{11}c_{(0,0)}  - u_{(1,1)}^{12}f_{(1,0)} ,\\
		z_{(1,1)x}=- 2\lambda z_{(1,1)}+ u_{(1,1)}^{8}h_{(0,0)} + u_{(1,1)}^{9}j_{(0,0)}-u_{(0,1)}^{10}e_{(1,0)}- u_{(0,1)}^{11}d_{(1,0)}+u_{(1,1)}^{12}b_{(0,0)} + u_{(1,1)}^{12}c_{(0,0)}.
	\end{cases}\label{r1}
\end{align}
Take the initial values
\begin{align*}
	a_{(0,0)}^{0}=\ b_{(0,0)}^{0}=c_{(0,0)}^{0}=1,\ d_{(1,0)}^{0}=e_{(1,0)}^{0}=\dots=w_{(0,1)}^{0}=z_{(1,1)}^{0}=\alpha^{0}(t)=0.
\end{align*}
By the recurrence relation(\ref{r1}), we have
\begin{align*}
	&a_{(0,0)}^{1}=\ b_{(0,0)}^{1}=c_{(0,0)}^{1}=\alpha^{1}(t)x,\ d_{(1,0)}^{1}=2\partial^{-1}(u_{(1,1)}^{1}u_{(0,1)}^{10}+u_{(0,1)}^{4}u_{(1,1)}^{8}+u_{(0,1)}^{5}u_{(1,1)}^{12}),\\ &e_{(1,0)}^{1}=2\partial^{-1}(u_{(1,1)}^{1}u_{(0,1)}^{11}+u_{(0,1)}^{4}u_{(1,1)}^{12}+u_{(0,1)}^{5}u_{(1,1)}^{9}),\ f_{(1,0)}^{1}=-2\partial^{-1}(u_{(1,1)}^{2}u_{(0,1)}^{10}+u_{(0,1)}^{4}u_{(1,1)}^{7}+u_{(1,1)}^{6}u_{(0,1)}^{11}),\\ 
	&g_{(1,0)}^{1}=-2\partial^{-1}(u_{(1,1)}^{3}u_{(0,1)}^{11}+u_{(0,1)}^{5}u_{(1,1)}^{7}+u_{(1,1)}^{6}u_{(0,1)}^{10}),\ h_{(0,0)}^{1}=0,\ j_{(0,0)}^{1}=0,\ k_{(1,1)}^{1}=u_{(1,1)}^{1},\ l_{(1,1)}^{1}=u_{(1,1)}^{2},\\
	&m_{(1,1)}^{1}=u_{(1,1)}^{3},\ n_{(0,1)}^{1}=u_{(0,1)}^{4},\ o_{(0,1)}^{1}=u_{(0,1)}^{5},\ p_{(1,1)}^{1}=u_{(1,1)}^{6},\ q_{(1,1)}^{1}=u_{(1,1)}^{7},\ r_{(1,1)}^{1}=u_{(1,1)}^{8},\ s_{(1,1)}^{1}=u_{(1,1)}^{9},\\
	&v_{(0,1)}^{1}=u_{(0,1)}^{10},\ w_{(0,1)}^{1}=u_{(0,1)}^{11},\ z_{(1,1)}^{1}=u_{(1,1)}^{12},\cdots
\end{align*}
Now, taking
\begin{align*}
	V^{n}=\lambda^{n}V=\sum_{i\geq0}(a_{(0,0)}^{i},\dots,z_{(1,1)}^{i})^{t}\lambda^{n-i},\quad V_{+}^{n}=\sum^{n}_{i=0}(a_{(0,0)}^{i},\dots,z_{(1,1)}^{i})^{t}\lambda^{n-i},\quad V_{-}^{n}=V^{n}-V_{+}^{n},
\end{align*}
according to the zero curvature equation $\frac{\partial U_{0}}{\partial u}u_{t}+\frac{\partial U_{0}}{\partial\lambda}\lambda_{t}-V_{0,+x}^{n}+[U_{0},V_{0,+}^{n}]=0$, we have
\allowdisplaybreaks
\begin{align*}
	u_{(1,1)t_{n}}^{1}=&\sum_{i=0}^{n} k_{(1,1)x}^{i}\lambda^{n-i}-2\lambda \sum_{i=0}^{n}k_{(1,1)}^{i}\lambda^{n-i}+ 2u_{(1,1)}^{1}\sum_{i=0}^{n} a_{(0,0)}^{i}\lambda^{n-i} + 2u_{(0,1)}^{4}\sum_{i=0}^{n} d_{(1,0)}^{i}\lambda^{n-i}\\& + 2u_{(0,1)}^{5}\sum_{i=0}^{n} e_{(1,0)}^{i}\lambda^{n-i} ,\\
	u_{(1,1)t_{n}}^{2}=&\sum_{i=0}^{n} l_{(1,1)x}^{i}\lambda^{n-i}-2\lambda \sum_{i=0}^{n} l_{(1,1)}^{i}\lambda^{n-i}+ 2u_{(1,1)}^{2}\sum_{i=0}^{n} b_{(0,0)}^{i}\lambda^{n-i} -2u_{(0,1)}^{4}\sum_{i=0}^{n} f_{(1,0)}^{i}\lambda^{n-i}\\& + 2u_{(1,1)}^{6}\sum_{i=0}^{n} h_{(0,0)}^{i}\lambda^{n-i} ,\\
	u_{(1,1)t_{n}}^{3}=&\sum_{i=0}^{n} m_{(1,1)x}^{i}\lambda^{n-i}-2\lambda \sum_{i=0}^{n} m_{(1,1)}^{i}\lambda^{n-i} + 2u_{(1,1)}^{3}\sum_{i=0}^{n} c_{(0,0)}^{i}\lambda^{n-i}-2u_{(0,1)}^{5}\sum_{i=0}^{n} g_{(1,0)}^{i}\lambda^{n-i}\\& + 2u_{(1,1)}^{6}\sum_{i=0}^{n} j_{(0,0)}^{i}\lambda^{n-i} ,\\
	u_{(0,1)t_{n}}^{4}=&\sum_{i=0}^{n} n_{(0,1)x}^{i}\lambda^{n-i}-2\lambda \sum_{i=0}^{n} n_{(0,1)}^{i}\lambda^{n-i} + u_{(1,1)}^{1}\sum_{i=0}^{n} f_{(1,0)}^{i}\lambda^{n-i}-u_{(1,1)}^{2}\sum_{i=0}^{n} d_{(1,0)}^{i}\lambda^{n-i}\\& + u_{(0,1)}^{4}\sum_{i=0}^{n} a_{(0,0)}^{i}\lambda^{n-i}  + u_{(0,1)}^{4}\sum_{i=0}^{n} b_{(0,0)}^{i}\lambda^{n-i}+ u_{(0,1)}^{5}\sum_{i=0}^{n} h_{(0,0)}^{i}\lambda^{n-i}- u_{(1,1)}^{6}\sum_{i=0}^{n} e_{(1,0)}^{i}\lambda^{n-i}  ,\\
	u_{(0,1)t_{n}}^{5}=&\sum_{i=0}^{n} o_{(0,1)x}^{i}\lambda^{n-i}-2\lambda \sum_{i=0}^{n} o_{(0,1)}^{i}\lambda^{n-i}+ u_{(1,1)}^{1}\sum_{i=0}^{n} g_{(1,0)}^{i}\lambda^{n-i}- u_{(1,1)}^{3}\sum_{i=0}^{n} e_{(1,0)}^{i}\lambda^{n-i}\\&+ u_{(0,1)}^{4}\sum_{i=0}^{n} j_{(0,0)}^{i}\lambda^{n-i}+ u_{(0,1)}^{5}\sum_{i=0}^{n} a_{(0,0)}^{i}\lambda^{n-i} + u_{(0,1)}^{5}\sum_{i=0}^{n} c_{(0,0)}^{i}\lambda^{n-i} -u_{(1,1)}^{6}\sum_{i=0}^{n} d_{(1,0)}^{i}\lambda^{n-i}   ,\\
	u_{(1,1)t_{n}}^{6}=&\sum_{i=0}^{n} p_{(1,1)x}^{i}\lambda^{n-i}-2\lambda \sum_{i=0}^{n} p_{(1,1)}^{i}\lambda^{n-i}+ u_{(1,1)}^{2}\sum_{i=0}^{n} j_{(0,0)}^{i}\lambda^{n-i}+ u_{(1,1)}^{3}\sum_{i=0}^{n} h_{(0,0)}^{i}\lambda^{n-i}\\&- u_{(0,1)}^{4}\sum_{i=0}^{n} g_{(1,0)}^{i}\lambda^{n-i} -u_{(0,1)}^{5}\sum_{i=0}^{n} f_{(1,0)}^{i}\lambda^{n-i} + u_{(1,1)}^{6}\sum_{i=0}^{n} b_{(0,0)}^{i}\lambda^{n-i}+u_{(1,1)}^{6}\sum_{i=0}^{n} c_{(0,0)}^{i}\lambda^{n-i},\\
	u_{(1,1)t_{n}}^{7}=&\sum_{i=0}^{n} q_{(1,1)x}^{i}\lambda^{n-i}+ 2\lambda \sum_{i=0}^{n} q_{(1,1)}^{i}\lambda^{n-i}-2u_{(1,1)}^{7}\sum_{i=0}^{n} a_{(0,0)}^{i}\lambda^{n-i} - 2u_{(0,1)}^{10}\sum_{i=0}^{n} f_{(1,0)}^{i}\lambda^{n-i}\\& - 2u_{(0,1)}^{11}\sum_{i=0}^{n} g_{(1,0)}^{i}\lambda^{n-i} ,\\
	u_{(1,1)t_{n}}^{8}=&\sum_{i=0}^{n} r_{(1,1)x}^{i}\lambda^{n-i}+ 2\lambda \sum_{i=0}^{n} r_{(1,1)}^{i}\lambda^{n-i}-2u_{(1,1)}^{8}\sum_{i=0}^{n} b_{(0,0)}^{i}\lambda^{n-i} + 2u_{(0,1)}^{10}\sum_{i=0}^{n} d_{(1,0)}^{i}\lambda^{n-i}\\& - 2u_{(1,1)}^{12}\sum_{i=0}^{n} j_{(0,0)}^{i}\lambda^{n-i} ,\\
	u_{(1,1)t_{n}}^{9}=&\sum_{i=0}^{n} s_{(1,1)x}^{i}\lambda^{n-i}+ 2\lambda \sum_{i=0}^{n} s_{(1,1)}^{i}\lambda^{n-i}-2u_{(1,1)}^{9}\sum_{i=0}^{n} c_{(0,0)}^{i}\lambda^{n-i} + 2u_{(0,1)}^{11}\sum_{i=0}^{n} e_{(1,0)}^{i}\lambda^{n-i}\\& - 2u_{(1,1)}^{12}\sum_{i=0}^{n} h_{(0,0)}^{i}\lambda^{n-i} ,\\
	u_{(0,1)t_{n}}^{10}=&\sum_{i=0}^{n} v_{(0,1)x}^{i}\lambda^{n-i}+ 2\lambda \sum_{i=0}^{n} v_{(0,1)}^{i}\lambda^{n-i}- u_{(1,1)}^{7}\sum_{i=0}^{n} d_{(1,0)}^{i}\lambda^{n-i} + u_{(1,1)}^{8}\sum_{i=0}^{n} f_{(1,0)}^{i}\lambda^{n-i}\\&-u_{(0,1)}^{10}\sum_{i=0}^{n} a_{(0,0)}^{i}\lambda^{n-i} - u_{(0,1)}^{10}\sum_{i=0}^{n} b_{(0,0)}^{i}\lambda^{n-i}-u_{(0,1)}^{11}\sum_{i=0}^{n} j_{(0,0)}^{i}\lambda^{n-i}+ u_{(1,1)}^{12}\sum_{i=0}^{n} g_{(1,0)}^{i}\lambda^{n-i},\\
	u_{(0,1)t_{n}}^{11}=&\sum_{i=0}^{n} w_{(0,1)x}^{i}\lambda^{n-i}+ 2\lambda \sum_{i=0}^{n} w_{(0,1)}^{i}\lambda^{n-i}-u_{(1,1)}^{7}\sum_{i=0}^{n} e_{(1,0)}^{i}\lambda^{n-i} + u_{(1,1)}^{9}\sum_{i=0}^{n} g_{(1,0)}^{i}\lambda^{n-i}\\& -u_{(0,1)}^{10}\sum_{i=0}^{n} h_{(0,0)}^{i}\lambda^{n-i}-u_{(0,1)}^{11}\sum_{i=0}^{n} a_{(0,0)}^{i}\lambda^{n-i} - u_{(0,1)}^{11}\sum_{i=0}^{n} c_{(0,0)}^{i}\lambda^{n-i}  + u_{(1,1)}^{12}\sum_{i=0}^{n} f_{(1,0)}^{i}\lambda^{n-i} ,\\
	u_{(1,1)t_{n}}^{12}=&\sum_{i=0}^{n} z_{(1,1)x}^{i}\lambda^{n-i}+ 2\lambda \sum_{i=0}^{n} z_{(1,1)}^{i}\lambda^{n-i}- u_{(1,1)}^{8}\sum_{i=0}^{n} h_{(0,0)}^{i}\lambda^{n-i} - u_{(1,1)}^{9}\sum_{i=0}^{n} j_{(0,0)}^{i}\lambda^{n-i}\\&+u_{(0,1)}^{10}\sum_{i=0}^{n} e_{(1,0)}^{i}\lambda^{n-i}+ u_{(0,1)}^{11}\sum_{i=0}^{n} d_{(1,0)}^{i}\lambda^{n-i}-u_{(1,1)}^{12}\sum_{i=0}^{n} b_{(0,0)}^{i}\lambda^{n-i} - u_{(1,1)}^{12}\sum_{i=0}^{n} c_{(0,0)}^{i}\lambda^{n-i}.
\end{align*}
This leads to the following Lax integrable hierarchy
\begin{align*}
	u_{t_{n}}=\left(
	\begin{array}{c}
		u_{(1,1)}^{1}\\
		u_{(1,1)}^{2}\\
		u_{(1,1)}^{3}\\
		u_{(0,1)}^{4}\\
		u_{(0,1)}^{5}\\
		u_{(1,1)}^{6}\\
		u_{(1,1)}^{7}\\
		u_{(1,1)}^{8}\\
		u_{(1,1)}^{9}\\
		u_{(0,1)}^{10}\\
		u_{(0,1)}^{11}\\
		u_{(1,1)}^{12}
	\end{array}\right)_{t_{n}}=\left(\begin{array}{c}
	2k_{(1,1)}^{n+1}\\2l_{(1,1)}^{n+1}\\2m_{(1,1)}^{n+1}\\2n_{(0,1)}^{n+1}\\2o_{(0,1)}^{n+1}\\2p_{(1,1)}^{n+1}\\-2q_{(1,1)}^{n+1}\\-2r_{(1,1)}^{n+1}\\-2s_{(1,1)}^{n+1}\\-2v_{(0,1)}^{n+1}\\-2w_{(0,1)}^{n+1}\\-2z_{(1,1)}^{n+1}\\
	\end{array}\right)\end{align*}\begin{align}
	=\left(\begin{array}{cccccccccccc}
		0&0&0&0&0&0&2&0&0&0&0&0\\
		0&0&0&0&0&0&0&2&0&0&0&0\\
		0&0&0&0&0&0&0&0&2&0&0&0\\
		0&0&0&0&0&0&0&0&0&-1&0&0\\
		0&0&0&0&0&0&0&0&0&0&-1&0\\
		0&0&0&0&0&0&0&0&0&0&0&1\\
		-2&0&0&0&0&0&0&0&0&0&0&0\\
		0&-2&0&0&0&0&0&0&0&0&0&0\\
		0&0&-2&0&0&0&0&0&0&0&0&0\\
		0&0&0&1&0&0&0&0&0&0&0&0\\
		0&0&0&0&1&0&0&0&0&0&0&0\\
		0&0&0&0&0&-1&0&0&0&0&0&0
	\end{array}\right)\left(\begin{array}{c}
	q_{(1,1)}^{n+1}\\r_{(1,1)}^{n+1}\\s_{(1,1)}^{n+1}\\-2v_{(0,1)}^{n+1}\\-2w_{(0,1)}^{n+1}\\2z_{(1,1)}^{n+1}\\k_{(1,1)}^{n+1}\\l_{(1,1)}^{n+1}\\m_{(1,1)}^{n+1}\\-2n_{(0,1)}^{n+1}\\-2o_{(0,1)}^{n+1}\\2p_{(1,1)}^{n+1}
	\end{array}\right)=JP^{n+1}.\label{h1}
\end{align}
From the recurrence relation (\ref{r1}), we have:
\begin{align*}
	P^{n+1}=LP^{n}+\bar{u}\alpha^{n}(t)x,
\end{align*}
where $L=(l_{i,j})_{12\times12}$ is detailed in \ref{app1}, and
\begin{align*}
	\bar{u}\alpha^{n}(t)x=(&u_{(1,1)}^{7}\alpha^{n}(t)x,u_{(1,1)}^{8}\alpha^{n}(t)x,u_{(1,1)}^{9}\alpha^{n}(t)x,-2u_{(0,1)}^{10}\alpha^{n}(t)x,-2u_{(0,1)}^{11}\alpha^{n}(t)x,2u_{(1,1)}^{12}\alpha^{n}(t)x,\\&u_{(1,1)}^{1}\alpha^{n}(t)x,u_{(1,1)}^{2}\alpha^{n}(t)x,u_{(1,1)}^{3}\alpha^{n}(t)x,-2_{(0,1)}u^{4}\alpha^{n}(t)x,-2u_{(0,1)}^{5}\alpha^{n}(t)x,-2u_{(1,1)}^{6}\alpha^{n}(t)x)^{t}.
\end{align*}
We derive the Hamiltonian structure of (\ref{h1}) via the trace identity,
\begin{align*}
	&\left\langle V,\frac{\partial U}{\partial\lambda}\right\rangle = 2a_{(0,0)}+2b_{(0,0)}+2c_{(0,0)} ,\
	\left\langle V,\frac{\partial U}{\partial u^{1}_{(1,1)}}\right\rangle = q_{(1,1)} ,\
	\left\langle V,\frac{\partial U}{\partial u^{2}_{(1,1)}}\right\rangle = r_{(1,1)} ,\\
	&\left\langle V,\frac{\partial U}{\partial u^{3}_{(1,1)}}\right\rangle = s_{(1,1)} ,\ 
	\left\langle V,\frac{\partial U}{\partial u^{4}_{(0,1)}}\right\rangle = -2v_{(0,1)} ,\
	\left\langle V,\frac{\partial U}{\partial u^{5}_{(0,1)}}\right\rangle = -2w_{(0,1)} ,\\
	&\left\langle V,\frac{\partial U}{\partial u^{6}_{(1,1)}}\right\rangle = 2z_{(1,1)} ,\
	\left\langle V,\frac{\partial U}{\partial u^{7}_{(1,1)}}\right\rangle = k_{(1,1)} ,\ 
	\left\langle V,\frac{\partial U}{\partial u^{8}_{(1,1)}}\right\rangle = l_{(1,1)} ,\ 
	\left\langle V,\frac{\partial U}{\partial u^{9}_{(1,1)}}\right\rangle = m_{(1,1)} ,\\
	&\left\langle V,\frac{\partial U}{\partial u^{10}_{(0,1)}}\right\rangle = -2n_{(0,1)} ,\
	\left\langle V,\frac{\partial U}{\partial u^{11}_{(0,1)}}\right\rangle = -2o_{(0,1)} ,\ 
	\left\langle V,\frac{\partial U}{\partial u^{12}_{(0,1)}}\right\rangle = 2p_{(1,1)} .\
\end{align*}
where $\left\langle X,Y\right\rangle=\mathrm{grtr}(XY)=\mathrm{tr}(XY)$ for all $X,Y\in\mathfrak{sp}_{1}(6)$. Substituting the above into the trace identity:
\begin{align*}
	\frac{\delta}{\delta u^{i}_{\mathbf{a}}}\left\langle V,\frac{\partial U}{\partial \lambda}\right\rangle=\lambda^{-\gamma}\frac{\partial}{\partial \lambda}\lambda^{\gamma}\left\langle V,\frac{\partial U}{\partial u^{i}_{\mathbf{a}}}\right\rangle
\end{align*}
 and balancing coefficients of each power of in the above equality gives rise to
 \begin{align*}
 	&\frac{\delta}{\delta u}(2a_{(0,0)}^{n+1}+2b_{(0,0)}^{n+1}+2c_{(0,0)}^{n+1})\\&=(\gamma-n)(q_{(1,1)}^{n},r_{(1,1)}^{n},s_{(1,1)}^{n},-2v_{(0,1)}^{n},-2w_{(0,1)}^{n},2z_{(1,1)}^{n},k_{(1,1)}^{n},l_{(1,1)}^{n},m_{(1,1)}^{n},-2n_{(0,1)}^{n},-2o_{(0,1)}^{n},2p_{(1,1)}^{n})^{t}.
 \end{align*}
 Setting $n=1$, we obtain $\gamma=0$ by direct computation. Therefore, for $P^{n+1}$ in equation (\ref{h1}), we have:
 \begin{align*}
 	P^{n+1}=\frac{\delta}{\delta u}(\frac{-2}{n+1}(a_{(0,0)}^{n+2}+b_{(0,0)}^{n+2}+c_{(0,0)}^{n+2})).
 \end{align*}
 Denoting $H^{n+1}=\frac{-2}{n+1}(a_{(0,0)}^{n+2}+b_{(0,0)}^{n+2}+c_{(0,0)}^{n+2})$, equation (2) can be expressed as:
 \begin{align*}
 	u_{t_{n}}=JP^{n+1}=J\frac{\delta}{\delta u}H^{n+1}=JLP^{n}+J\bar{u}\alpha^{n}(t)x=JL\frac{\delta}{\delta u}H^{n}+J\bar{u}\alpha^{n}(t)x.
 \end{align*}
 Thus, we have constructed a new class of super‑integrable hierarchies together with their super‑Hamiltonian structures on the color Lie algebra $\mathfrak{sp}_{1}(6)$.

 \section{The $(2+1)$-dimensional non-isospectral integrable hierarchy associated with color Lie algebra $\mathfrak{sp}_{1}(6)$}\label{sec3}
 Take
 \begin{align*}
 	\frac{\partial}{\partial X}=\frac{\partial}{\partial y}-\frac{\partial}{\partial x},\ \frac{\partial}{\partial Y}=\frac{\partial}{\partial t}-\frac{\partial}{\partial x},
 \end{align*}
 we consider a non-isospectral problem
 \begin{align*}
 	\begin{cases}
 		\phi_{X}=U\phi\\
 		\phi_{Y}=V\phi\\
 		\lambda_{t}=\sum_{i\geq0}\alpha^{i}(t)\lambda^{-i}
 	\end{cases}
 \end{align*}
where $U$ and $V$ are given by equations (\ref{u}) and (\ref{v}), respectively. The zero curvature equation $U_{Y}-V_{X}+[U,V]=0$ can be rewritten in (2+1)-dimensional form $U_{t}-U_{x}+V_{x}-V_{y}+[U,V]=0$, the stationary zero curvature equation $V_{X}=\frac{\partial U}{\partial\lambda}\lambda_{t}+[U,V]$ can be rewritten as $V_{y}-V_{x}=\frac{\partial U}{\partial\lambda}\lambda_{t}+[U,V]$, and satisfies the compatibility condition for the following Lax pair
\begin{align*}
	\begin{cases}
		\phi_{y}=\phi_{x}+U\phi\\
		\phi_{t}=\phi_{x}+V\phi
	\end{cases}
\end{align*}
The stationary zero curvature equation $V_{y}-V_{x}=\frac{\partial U}{\partial\lambda}\lambda_{t}+[U,V]$ gives
\begin{align}
	\begin{cases}
		a_{(0,0)x}=a_{(0,0)y}-u_{(1,1)}^{1}q_{(1,1)} + u_{(0,1)}^{4}v_{(0,1)} + u_{(0,1)}^{5}w_{(0,1)}+ u_{(1,1)}^{7}k_{(1,1)} - u_{(0,1)}^{10}n_{(0,1)} - u_{(0,1)}^{11}o_{(0,1)}  -\lambda_{t} ,\\
		b_{(0,0)x}=b_{(0,0)y}-u_{(1,1)}^{2}r_{(1,1)} + u_{(0,1)}^{4}v_{(0,1)} - u_{(1,1)}^{6}z_{(1,1)}+ u_{(1,1)}^{8}l_{(1,1)} - u_{(0,1)}^{10}n_{(0,1)} + u_{(1,1)}^{12}p_{(1,1)} -\lambda_{t} ,\\
		c_{(0,0)x}=c_{(0,0)y}-u_{(1,1)}^{3}s_{(1,1)} + u_{(0,1)}^{5}w_{(0,1)} - u_{(1,1)}^{6}z_{(1,1)}+ u_{(1,1)}^{9}m_{(1,1)} - u_{(0,1)}^{11}o_{(0,1)} + u_{(1,1)}^{12}p_{(1,1)}-\lambda_{t} ,\\
		d_{(1,0)x}=d_{(1,0)y}-u_{(1,1)}^{1}v_{(0,1)}-u_{(0,1)}^{4}r_{(1,1)} - u_{(0,1)}^{5}z_{(1,1)} - u_{(1,1)}^{8}n_{(0,1)} -u_{(0,1)}^{10}k_{(1,1)}- u_{(1,1)}^{12}o_{(0,1)},\\
		e_{(1,0)x}=e_{(1,0)y}-u_{(1,1)}^{1}w_{(0,1)} -u_{(0,1)}^{4}z_{(1,1)}-u_{(0,1)}^{5}s_{(1,1)}-u_{(1,1)}^{9}o_{(0,1)}-u_{(0,1)}^{11}k_{(1,1)} - u_{(1,1)}^{12}n_{(0,1)} ,\\
		f_{(1,0)x}=f_{(1,0)y}+u_{(1,1)}^{2}v_{(0,1)}+u_{(0,1)}^{4}q_{(1,1)}+ u_{(1,1)}^{6}w_{(0,1)}+ u_{(1,1)}^{7}n_{(0,1)}+ u_{(0,1)}^{10}l_{(1,1)}  + u_{(0,1)}^{11}p_{(1,1)},\\
		g_{(1,0)x}=g_{(1,0)y}+u_{(1,1)}^{3}w_{(0,1)}+u_{(0,1)}^{5}q_{(1,1)} +u_{(1,1)}^{6}v_{(0,1)}+u_{(1,1)}^{7}o_{(0,1)}+ u_{(0,1)}^{10}p_{(1,1)}+u_{(0,1)}^{11}m_{(1,1)}   ,\\
		h_{(0,0)x}=h_{(0,0)y}-u_{(1,1)}^{2}z_{(1,1)}+u_{(0,1)}^{4}w_{(0,1)}- u_{(1,1)}^{6}s_{(1,1)}+u_{(1,1)}^{9}p_{(1,1)}-u_{(0,1)}^{11}n_{(0,1)}+u_{(1,1)}^{12}l_{(1,1)},\\
		j_{(0,0)x}=j_{(0,0)y}-u_{(1,1)}^{3}z_{(1,1)}+u_{(0,1)}^{5}v_{(0,1)} - u_{(1,1)}^{6}r_{(1,1)}+u_{(1,1)}^{8}p_{(1,1)}-u_{(0,1)}^{10}o_{(0,1)}+u_{(1,1)}^{12}m_{(1,1)},\\
		k_{(1,1)x}=k_{(1,1)y}-2\lambda k_{(1,1)}+ 2u_{(1,1)}^{1}a_{(0,0)} + 2u_{(0,1)}^{4}d_{(1,0)} + 2u_{(0,1)}^{5}e_{(1,0)} ,\\
		l_{(1,1)x}=l_{(1,1)y}-2\lambda l_{(1,1)}+ 2u_{(1,1)}^{2}b_{(0,0)} -2u_{(0,1)}^{4}f_{(1,0)} + 2u_{(1,1)}^{6}h_{(0,0)} ,\\
		m_{(1,1)x}=m_{(1,1)y}-2\lambda m_{(1,1)} + 2u_{(1,1)}^{3}c_{(0,0)}-2u_{(0,1)}^{5}g_{(1,0)} + 2u_{(1,1)}^{6}j_{(0,0)} ,\\
		n_{(0,1)x}=n_{(0,1)y}-2\lambda n_{(0,1)} + u_{(1,1)}^{1}f_{(1,0)}-u_{(1,1)}^{2}d_{(1,0)} + u_{(0,1)}^{4}a_{(0,0)}  + u_{(0,1)}^{4}b_{(0,0)}+ u_{(0,1)}^{5}h_{(0,0)}- u_{(1,1)}^{6}e_{(1,0)}  ,\\
		o_{(0,1)x}=o_{(0,1)y}-2\lambda o_{(0,1)}+ u_{(1,1)}^{1}g_{(1,0)}- u_{(1,1)}^{3}e_{(1,0)}+ u_{(0,1)}^{4}j_{(0,0)}+ u_{(0,1)}^{5}a_{(0,0)} + u_{(0,1)}^{5}c_{(0,0)} -u_{(1,1)}^{6}d_{(1,0)}   ,\\
		p_{(1,1)x}=p_{(1,1)y}-2\lambda p_{(1,1)}+ u_{(1,1)}^{2}j_{(0,0)}+ u_{(1,1)}^{3}h_{(0,0)}- u_{(0,1)}^{4}g_{(1,0)} -u_{(0,1)}^{5}f_{(1,0)} + u_{(1,1)}^{6}b_{(0,0)}+u_{(1,1)}^{6}c_{(0,0)},\\
		q_{(1,1)x}=q_{(1,1)y}+ 2\lambda q_{(1,1)}-2u_{(1,1)}^{7}a_{(0,0)} - 2u_{(0,1)}^{10}f_{(1,0)} - 2u_{(0,1)}^{11}g_{(1,0)} ,\\
		r_{(1,1)x}=r_{(1,1)y}+ 2\lambda r_{(1,1)}-2u_{(1,1)}^{8}b_{(0,0)} + 2u_{(0,1)}^{10}d_{(1,0)} - 2u_{(1,1)}^{12}j_{(0,0)} ,\\
		s_{(1,1)x}=s_{(1,1)y}+ 2\lambda s_{(1,1)}-2u_{(1,1)}^{9}c_{(0,0)} + 2u_{(0,1)}^{11}e_{(1,0)} - 2u_{(1,1)}^{12}h_{(0,0)} ,\\
		v_{(0,1)x}=v_{(0,1)y}+ 2\lambda v_{(0,1)}- u_{(1,1)}^{7}d_{(1,0)} + u_{(1,1)}^{8}f_{(1,0)}-u_{(0,1)}^{10}a_{(0,0)} - u_{(0,1)}^{10}b_{(0,0)}-u_{(0,1)}^{11}j_{(0,0)}+ u_{(1,1)}^{12}g_{(1,0)},\\
		w_{(0,1)x}=w_{(0,1)y}+ 2\lambda w_{(0,1)}-u_{(1,1)}^{7}e_{(1,0)} + u_{(1,1)}^{9}g_{(1,0)} -u_{(0,1)}^{10}h_{(0,0)}-u_{(0,1)}^{11}a_{(0,0)} - u_{(0,1)}^{11}c_{(0,0)}  + u_{(1,1)}^{12}f_{(1,0)} ,\\
		z_{(1,1)x}=z_{(1,1)y}+ 2\lambda z_{(1,1)}- u_{(1,1)}^{8}h_{(0,0)} - u_{(1,1)}^{9}j_{(0,0)}+u_{(0,1)}^{10}e_{(1,0)}+ u_{(0,1)}^{11}d_{(1,0)}-u_{(1,1)}^{12}b_{(0,0)} - u_{(1,1)}^{12}c_{(0,0)}.
	\end{cases}\label{r2}
\end{align}
We can compute the initial terms of the recurrence relation under the initial values specified in Section \ref{sec2}:
\begin{align*}
	&a_{(0,0)}^{1}=\partial^{-1}(a_{(0,0)y}^{1}-\alpha^{1}(t)x),\  b_{(0,0)}^{1}=\partial^{-1}(b_{(0,0)y}^{1}-\alpha^{1}(t)x),\  c_{(0,0)}^{1}=\partial^{-1}(c_{(0,0)y}^{1}-\alpha^{1}(t)x),\\
	&d_{(1,0)}^{1}=-2\partial^{-1}(-\frac{1}{2}d_{(1,0)y}^{1}+u_{(1,1)}^{1}u_{(0,1)}^{10}+u_{(0,1)}^{4}u_{(1,1)}^{8}+u_{(0,1)}^{5}u_{(1,1)}^{12}),\\
	&e_{(1,0)}^{1}=-2\partial^{-1}(-\frac{1}{2}e_{(1,0)y}^{1}+u_{(1,1)}^{1}u_{(0,1)}^{11}+u_{(0,1)}^{4}u_{(1,1)}^{12}+u_{(0,1)}^{5}u_{(1,1)}^{9}),\\
	&f_{(1,0)}^{1}=2\partial^{-1}(\frac{1}{2}f_{(1,0)y}^{1}+u_{(1,1)}^{2}u_{(0,1)}^{10}+u_{(0,1)}^{4}u_{(1,1)}^{7}+u_{(1,1)}^{6}u_{(0,1)}^{11}),\\ 
	&g_{(1,0)}^{1}=2\partial^{-1}(\frac{1}{2}g_{(1,0)y}^{1}+u_{(1,1)}^{3}u_{(0,1)}^{11}+u_{(0,1)}^{5}u_{(1,1)}^{7}+u_{(1,1)}^{6}u_{(0,1)}^{10}),\\
	&h_{(0,0)}^{1}=\partial^{-1}h_{(0,0)y}^{1},\ j_{(0,0)}^{1}=\partial^{-1}j_{(0,0)y}^{1},\ k_{(1,1)}^{1}=u_{(1,1)}^{1},\ l_{(1,1)}^{1}=u_{(1,1)}^{2},\\
	&m_{(1,1)}^{1}=u_{(1,1)}^{3},\ n_{(0,1)}^{1}=u_{(0,1)}^{4},\ o_{(0,1)}^{1}=u_{(0,1)}^{5},\ p_{(1,1)}^{1}=u_{(1,1)}^{6},\ q_{(1,1)}^{1}=u_{(1,1)}^{7},\\ 
	&r_{(1,1)}^{1}=u_{(1,1)}^{8},\ s_{(1,1)}^{1}=u_{(1,1)}^{9},\ v_{(0,1)}^{1}=u_{(0,1)}^{10},\ w_{(0,1)}^{1}=u_{(0,1)}^{11},\ z_{(1,1)}^{1}=u_{(1,1)}^{12},\cdots
\end{align*}
Now, taking
\begin{align*}
	V^{n}=\lambda^{n}V=\sum_{i\geq0}(a_{(0,0)}^{i},\dots,z_{(1,1)}^{i})^{t}\lambda^{n-i},\quad V_{+}^{n}=\sum^{n}_{i=0}(a_{(0,0)}^{i},\dots,z_{(1,1)}^{i})^{t}\lambda^{n-i},\quad V_{-}^{n}=V^{n}-V_{+}^{n},
\end{align*}
According to the zero curvature equation
$U_{t}-U_{x}+V_{+x}^{(n)}-V_{+y}^{(n)}+[U,V_{+}^{(n)}]=0$, we have
\allowdisplaybreaks
\begin{align*}
	u_{(1,1)t_{n}}^{1}=&-\sum_{i=0}^{n}k_{(1,1)x}^{i}\lambda^{n-i}+\sum_{i=0}^{n}k_{(1,1)y}^{i}\lambda^{n-i}-2\lambda \sum_{i=0}^{n}k_{(1,1)}^{i}\lambda^{n-i}+ 2u_{(1,1)}^{1}\sum_{i=0}^{n} a_{(0,0)}^{i}\lambda^{n-i}\\& + 2u_{(0,1)}^{4}\sum_{i=0}^{n} d_{(1,0)}^{i}\lambda^{n-i}+ 2u_{(0,1)}^{5}\sum_{i=0}^{n} e_{(1,0)}^{i}\lambda^{n-i}+u_{(1,1)x}^{1} ,\\
	u_{(1,1)t_{n}}^{2}=&-\sum_{i=0}^{n} l_{(1,1)x}^{i}\lambda^{n-i}+\sum_{i=0}^{n} l_{(1,1)y}^{i}\lambda^{n-i}-2\lambda \sum_{i=0}^{n} l_{(1,1)}^{i}\lambda^{n-i}+ 2u_{(1,1)}^{2}\sum_{i=0}^{n} b_{(0,0)}^{i}\lambda^{n-i}\\& -2u_{(0,1)}^{4}\sum_{i=0}^{n} f_{(1,0)}^{i}\lambda^{n-i} + 2u_{(1,1)}^{6}\sum_{i=0}^{n} h_{(0,0)}^{i}\lambda^{n-i}+u_{(1,1)x}^{2} ,\\
	u_{(1,1)t_{n}}^{3}=&-\sum_{i=0}^{n} m_{(1,1)x}^{i}\lambda^{n-i}+\sum_{i=0}^{n} m_{(1,1)y}^{i}\lambda^{n-i}-2\lambda \sum_{i=0}^{n} m_{(1,1)}^{i}\lambda^{n-i} + 2u_{(1,1)}^{3}\sum_{i=0}^{n} c_{(0,0)}^{i}\lambda^{n-i}\\&-2u_{(0,1)}^{5}\sum_{i=0}^{n} g_{(1,0)}^{i}\lambda^{n-i} + 2u_{(1,1)}^{6}\sum_{i=0}^{n} j_{(0,0)}^{i}\lambda^{n-i}+u_{(1,1)x}^{3} ,\\
	u_{(0,1)t_{n}}^{4}=&-\sum_{i=0}^{n} n_{(0,1)x}^{i}\lambda^{n-i}+\sum_{i=0}^{n} n_{(0,1)y}^{i}\lambda^{n-i}-2\lambda \sum_{i=0}^{n} n_{(0,1)}^{i}\lambda^{n-i} + u_{(1,1)}^{1}\sum_{i=0}^{n} f_{(1,0)}^{i}\lambda^{n-i}\\&-u_{(1,1)}^{2}\sum_{i=0}^{n} d_{(1,0)}^{i}\lambda^{n-i} + u_{(0,1)}^{4}\sum_{i=0}^{n} a_{(0,0)}^{i}\lambda^{n-i}  + u_{(0,1)}^{4}\sum_{i=0}^{n} b_{(0,0)}^{i}\lambda^{n-i}\\&+ u_{(0,1)}^{5}\sum_{i=0}^{n} h_{(0,0)}^{i}\lambda^{n-i}- u_{(1,1)}^{6}\sum_{i=0}^{n} e_{(1,0)}^{i}\lambda^{n-i}+u_{(0,1)x}^{4}  ,\\
	u_{(0,1)t_{n}}^{5}=&-\sum_{i=0}^{n} o_{(0,1)x}^{i}\lambda^{n-i}+\sum_{i=0}^{n} o_{(0,1)y}^{i}\lambda^{n-i}-2\lambda \sum_{i=0}^{n} o_{(0,1)}^{i}\lambda^{n-i}+ u_{(1,1)}^{1}\sum_{i=0}^{n} g_{(1,0)}^{i}\lambda^{n-i}\\&- u_{(1,1)}^{3}\sum_{i=0}^{n} e_{(1,0)}^{i}\lambda^{n-i}+ u_{(0,1)}^{4}\sum_{i=0}^{n} j_{(0,0)}^{i}\lambda^{n-i}+ u_{(0,1)}^{5}\sum_{i=0}^{n} a_{(0,0)}^{i}\lambda^{n-i} + u_{(0,1)}^{5}\sum_{i=0}^{n} c_{(0,0)}^{i}\lambda^{n-i}\\& -u_{(1,1)}^{6}\sum_{i=0}^{n} d_{(1,0)}^{i}\lambda^{n-i}+u_{(0,1)x}^{5}   ,\\
	u_{(1,1)t_{n}}^{6}=&-\sum_{i=0}^{n} p_{(1,1)x}^{i}\lambda^{n-i}+\sum_{i=0}^{n} p_{(1,1)y}^{i}\lambda^{n-i}-2\lambda \sum_{i=0}^{n} p_{(1,1)}^{i}\lambda^{n-i}+ u_{(1,1)}^{2}\sum_{i=0}^{n} j_{(0,0)}^{i}\lambda^{n-i}\\&+ u_{(1,1)}^{3}\sum_{i=0}^{n} h_{(0,0)}^{i}\lambda^{n-i}- u_{(0,1)}^{4}\sum_{i=0}^{n} g_{(1,0)}^{i}\lambda^{n-i} -u_{(0,1)}^{5}\sum_{i=0}^{n} f_{(1,0)}^{i}\lambda^{n-i} + u_{(1,1)}^{6}\sum_{i=0}^{n} b_{(0,0)}^{i}\lambda^{n-i}\\&+u_{(1,1)}^{6}\sum_{i=0}^{n} c_{(0,0)}^{i}\lambda^{n-i}+u_{(1,1)x}^{6},\\
	u_{(1,1)t_{n}}^{7}=&-\sum_{i=0}^{n} q_{(1,1)x}^{i}\lambda^{n-i}+\sum_{i=0}^{n} q_{(1,1)y}^{i}\lambda^{n-i}+ 2\lambda \sum_{i=0}^{n} q_{(1,1)}^{i}\lambda^{n-i}-2u_{(1,1)}^{7}\sum_{i=0}^{n} a_{(0,0)}^{i}\lambda^{n-i}\\& - 2u_{(0,1)}^{10}\sum_{i=0}^{n} f_{(1,0)}^{i}\lambda^{n-i} - 2u_{(0,1)}^{11}\sum_{i=0}^{n} g_{(1,0)}^{i}\lambda^{n-i}+u_{(1,1)x}^{7} ,\\
	u_{(1,1)t_{n}}^{8}=&-\sum_{i=0}^{n} r_{(1,1)x}^{i}\lambda^{n-i}+\sum_{i=0}^{n} r_{(1,1)y}^{i}\lambda^{n-i}+ 2\lambda \sum_{i=0}^{n} r_{(1,1)}^{i}\lambda^{n-i}-2u_{(1,1)}^{8}\sum_{i=0}^{n} b_{(0,0)}^{i}\lambda^{n-i}\\& + 2u_{(0,1)}^{10}\sum_{i=0}^{n} d_{(1,0)}^{i}\lambda^{n-i} - 2u_{(1,1)}^{12}\sum_{i=0}^{n} j_{(0,0)}^{i}\lambda^{n-i}u_{(1,1)x}^{8} ,\\
	u_{(1,1)t_{n}}^{9}=&-\sum_{i=0}^{n} s_{(1,1)x}^{i}\lambda^{n-i}+\sum_{i=0}^{n} s_{(1,1)y}^{i}\lambda^{n-i}+ 2\lambda \sum_{i=0}^{n} s_{(1,1)}^{i}\lambda^{n-i}-2u_{(1,1)}^{9}\sum_{i=0}^{n} c_{(0,0)}^{i}\lambda^{n-i}\\& + 2u_{(0,1)}^{11}\sum_{i=0}^{n} e_{(1,0)}^{i}\lambda^{n-i} - 2u_{(1,1)}^{12}\sum_{i=0}^{n} h_{(0,0)}^{i}\lambda^{n-i}u_{(1,1)x}^{9} ,\\
	u_{(0,1)t_{n}}^{10}=&-\sum_{i=0}^{n} v_{(0,1)x}^{i}\lambda^{n-i}+\sum_{i=0}^{n} v_{(0,1)y}^{i}\lambda^{n-i}+ 2\lambda \sum_{i=0}^{n} v_{(0,1)}^{i}\lambda^{n-i}- u_{(1,1)}^{7}\sum_{i=0}^{n} d_{(1,0)}^{i}\lambda^{n-i}\\& + u_{(1,1)}^{8}\sum_{i=0}^{n} f_{(1,0)}^{i}\lambda^{n-i}-u_{(0,1)}^{10}\sum_{i=0}^{n} a_{(0,0)}^{i}\lambda^{n-i} - u_{(0,1)}^{10}\sum_{i=0}^{n} b_{(0,0)}^{i}\lambda^{n-i}-u_{(0,1)}^{11}\sum_{i=0}^{n} j_{(0,0)}^{i}\lambda^{n-i}\\&+ u_{(1,1)}^{12}\sum_{i=0}^{n} g_{(1,0)}^{i}\lambda^{n-i}+u_{(0,1)x}^{10},\\
	u_{(0,1)t_{n}}^{11}=&-\sum_{i=0}^{n} w_{(0,1)x}^{i}\lambda^{n-i}+\sum_{i=0}^{n} w_{(0,1)y}^{i}\lambda^{n-i}+ 2\lambda \sum_{i=0}^{n} w_{(0,1)}^{i}\lambda^{n-i}-u_{(1,1)}^{7}\sum_{i=0}^{n} e_{(1,0)}^{i}\lambda^{n-i}\\& + u_{(1,1)}^{9}\sum_{i=0}^{n} g_{(1,0)}^{i}\lambda^{n-i} -u_{(0,1)}^{10}\sum_{i=0}^{n} h_{(0,0)}^{i}\lambda^{n-i}-u_{(0,1)}^{11}\sum_{i=0}^{n} a_{(0,0)}^{i}\lambda^{n-i} - u_{(0,1)}^{11}\sum_{i=0}^{n} c_{(0,0)}^{i}\lambda^{n-i}  \\&+ u_{(1,1)}^{12}\sum_{i=0}^{n} f_{(1,0)}^{i}\lambda^{n-i}+u_{(0,1)x}^{11} ,\\
	u_{(1,1)t_{n}}^{12}=&-\sum_{i=0}^{n} z_{(1,1)x}^{i}\lambda^{n-i}+\sum_{i=0}^{n} z_{(1,1)y}^{i}\lambda^{n-i}+ 2\lambda \sum_{i=0}^{n} z_{(1,1)}^{i}\lambda^{n-i}- u_{(1,1)}^{8}\sum_{i=0}^{n} h_{(0,0)}^{i}\lambda^{n-i}\\& - u_{(1,1)}^{9}\sum_{i=0}^{n} j_{(0,0)}^{i}\lambda^{n-i}+u_{(0,1)}^{10}\sum_{i=0}^{n} e_{(1,0)}^{i}\lambda^{n-i}+ u_{(0,1)}^{11}\sum_{i=0}^{n} d_{(1,0)}^{i}\lambda^{n-i}-u_{(1,1)}^{12}\sum_{i=0}^{n} b_{(0,0)}^{i}\lambda^{n-i} \\&- u_{(1,1)}^{12}\sum_{i=0}^{n} c_{(0,0)}^{i}\lambda^{n-i}+u_{(1,1)x}^{12}.
\end{align*}
This leads to the following Lax integrable hierarchy
\begin{align*}
	u_{t_{n}}=\left(
	\begin{array}{c}
		u_{(1,1)}^{1}\\
		u_{(1,1)}^{2}\\
		u_{(1,1)}^{3}\\
		u_{(0,1)}^{4}\\
		u_{(0,1)}^{5}\\
		u_{(1,1)}^{6}\\
		u_{(1,1)}^{7}\\
		u_{(1,1)}^{8}\\
		u_{(1,1)}^{9}\\
		u_{(0,1)}^{10}\\
		u_{(0,1)}^{11}\\
		u_{(1,1)}^{12}
	\end{array}\right)_{t_{n}}=\left(\begin{array}{c}
		2k_{(1,1)}^{n+1}\\2l_{(1,1)}^{n+1}\\2m_{(1,1)}^{n+1}\\2n_{(0,1)}^{n+1}\\2o_{(0,1)}^{n+1}\\2p_{(1,1)}^{n+1}\\-2q_{(1,1)}^{n+1}\\-2r_{(1,1)}^{n+1}\\-2s_{(1,1)}^{n+1}\\-2v_{(0,1)}^{n+1}\\-2w_{(0,1)}^{n+1}\\-2z_{(1,1)}^{n+1}
	\end{array}\right)+\left(\begin{array}{c}
	u_{(1,1)x}^{1}\\u_{(1,1)x}^{2}\\u_{(1,1)x}^{3}\\u_{(0,1)x}^{4}\\u_{(0,1)x}^{5}\\u_{(1,1)x}^{6}\\u_{(1,1)x}^{7}\\u_{(1,1)x}^{8}\\u_{(1,1)x}^{9}\\u_{(0,1)x}^{10}\\u_{(0,1)x}^{11}\\u_{(1,1)x}^{12}
	\end{array}\right)\end{align*}\begin{align}
	=\left(\begin{array}{cccccccccccc}
	0&0&0&0&0&0&2&0&0&0&0&0\\
	0&0&0&0&0&0&0&2&0&0&0&0\\
	0&0&0&0&0&0&0&0&2&0&0&0\\
	0&0&0&0&0&0&0&0&0&-1&0&0\\
	0&0&0&0&0&0&0&0&0&0&-1&0\\
	0&0&0&0&0&0&0&0&0&0&0&1\\
	-2&0&0&0&0&0&0&0&0&0&0&0\\
	0&-2&0&0&0&0&0&0&0&0&0&0\\
	0&0&-2&0&0&0&0&0&0&0&0&0\\
	0&0&0&1&0&0&0&0&0&0&0&0\\
	0&0&0&0&1&0&0&0&0&0&0&0\\
	0&0&0&0&0&-1&0&0&0&0&0&0
	\end{array}\right)\left(\begin{array}{c}
	q_{(1,1)}^{n+1}\\r_{(1,1)}^{n+1}\\s_{(1,1)}^{n+1}\\-2v_{(0,1)}^{n+1}\\-2w_{(0,1)}^{n+1}\\2z_{(1,1)}^{n+1}\\k_{(1,1)}^{n+1}\\l_{(1,1)}^{n+1}\\m_{(1,1)}^{n+1}\\-2n_{(0,1)}^{n+1}\\-2o_{(0,1)}^{n+1}\\2p_{(1,1)}^{n+1}
	\end{array}\right)+\left(\begin{array}{c}
	u_{(1,1)x}^{1}\\u_{(1,1)x}^{2}\\u_{(1,1)x}^{3}\\u_{(0,1)x}^{4}\\u_{(0,1)x}^{5}\\u_{(1,1)x}^{6}\\u_{(1,1)x}^{7}\\u_{(1,1)x}^{8}\\u_{(1,1)x}^{9}\\u_{(0,1)x}^{10}\\u_{(0,1)x}^{11}\\u_{(1,1)x}^{12}
	\end{array}\right)=JP^{n+1}+u_{x}.\label{h2}
\end{align}
From the recurrence relations (\ref{r2}), we have
\begin{align*}
	P^{n+1}=-LP^{n}-\left(\begin{array}{c}
		u_{(1,1)}^{7}\alpha^{n}(t)x\\u_{(1,1)}^{8}\alpha^{n}(t)x\\u_{(1,1)}^{9}\alpha^{n}(t)x\\-2u_{(0,1)}^{10}\alpha^{n}(t)x\\-2u_{(0,1)}^{11}\alpha^{n}(t)x\\2u_{(1,1)}^{12}\alpha^{n}(t)x\\u_{(1,1)}^{1}\alpha^{n}(t)x\\u_{(1,1)}^{2}\alpha^{n}(t)x\\u_{(1,1)}^{3}\alpha^{n}(t)x\\-2u_{(0,1)}^{4}\alpha^{n}(t)x\\-2u_{(0,1)}^{5}\alpha^{n}(t)x\\2u_{(1,1)}^{6}\alpha^{n}(t)x
	\end{array}\right)
\end{align*}
\scalebox{0.85}{%
$+\left(\begin{array}{c}
		-\frac{1}{2}q_{(1,1)y}^{n}+u_{(1,1)}^{7}\partial^{-1}a_{(0,0)y}^{n}+u_{(0,1)}^{10}\partial^{-1}f_{(1,0)y}^{n}+u_{(0,1)}^{11}\partial^{-1}g_{(1,0)y}^{n}\\
		-\frac{1}{2}r_{(1,1)y}^{n}+u_{(1,1)}^{8}\partial^{-1}b_{(0,0)y}^{n}-u_{(0,1)}^{10}\partial^{-1}d_{(1,0)y}^{n}+u_{(1,1)}^{12}\partial^{-1}j_{(0,0)y}^{n}\\
		-\frac{1}{2}s_{(1,1)y}^{n}+u_{(1,1)}^{9}\partial^{-1}c_{(0,0)y}^{n}-u_{(0,1)}^{11}\partial^{-1}e_{(1,0)y}^{n}+u_{(1,1)}^{12}\partial^{-1}h_{(0,0)y}^{n}\\
		v_{(0,1)y}^{n}- u_{(1,1)}^{7}\partial^{-1}d_{(1,0)y}^{n} + u_{(1,1)}^{8}\partial^{-1}f_{(1,0)y}^{n}-u_{(0,1)}^{10}\partial^{-1}a_{(0,0)y}^{n} - u_{(0,1)}^{10}\partial^{-1}b_{(0,0)y}^{n}-u_{(0,1)}^{11}\partial^{-1}j_{(0,0)y}^{n}+ u_{(1,1)}^{12}\partial^{-1}g_{(1,0)y}^{n}\\
		w_{(0,1)y}^{n}-u_{(1,1)}^{7}\partial^{-1}e_{(1,0)y}^{n} + u_{(1,1)}^{9}\partial^{-1}g_{(1,0)y}^{n} -u_{(0,1)}^{10}\partial^{-1}h_{(0,0)y}^{n}-u_{(0,1)}^{11}\partial^{-1}a_{(0,0)y}^{n} - u_{(0,1)}^{11}\partial^{-1}c_{(0,0)y}^{n}  + u_{(1,1)}^{12}\partial^{-1}f_{(1,0)y}^{n} \\
		-z_{(1,1)y}^{n}+ u_{(1,1)}^{8}\partial^{-1}h_{(0,0)y}^{n} + u_{(1,1)}^{9}\partial^{-1}j_{(0,0)y}^{n}-u_{(0,1)}^{10}\partial^{-1}e_{(1,0)y}^{n}- u_{(0,1)}^{11}\partial^{-1}d_{(1,0)y}^{n}+u_{(1,1)}^{12}\partial^{-1}b_{(0,0)y}^{n} + u_{(1,1)}^{12}\partial^{-1}c_{(0,0)y}^{n}\\
		\frac{1}{2}k_{(1,1)y}^{n}+ u_{(1,1)}^{1}\partial^{-1}a_{(0,0)y}^{n} + u_{(0,1)}^{4}\partial^{-1}d_{(1,0)y}^{n} + u_{(0,1)}^{5}\partial^{-1}e_{(1,0)y}^{n}\\
		\frac{1}{2}l_{(1,1)y}^{n}+ u_{(1,1)}^{2}\partial^{-1}b_{(0,0)y}^{n} -u_{(0,1)}^{4}\partial^{-1}f_{(1,0)y}^{n} + u_{(1,1)}^{6}\partial^{-1}h_{(0,0)y}^{n}\\
		\frac{1}{2}m_{(1,1)y}^{n}+ u_{(1,1)}^{3}\partial^{-1}c_{(0,0)y}^{n}-u_{(0,1)}^{5}\partial^{-1}g_{(1,0)y}^{n} + u_{(1,1)}^{6}\partial^{-1}j_{(0,0)y}^{n}\\
		-n_{(0,1)y}^{n}- u_{(1,1)}^{1}\partial^{-1}f_{(1,0)y}^{n}+u_{(1,1)}^{2}\partial^{-1}d_{(1,0)y}^{n} - u_{(0,1)}^{4}\partial^{-1}a_{(0,0)y}^{n}-u_{(0,1)}^{4}\partial^{-1}b_{(0,0)y}^{n}- u_{(0,1)}^{5}\partial^{-1}h_{(0,0)y}^{n}+ u_{(1,1)}^{6}\partial^{-1}e_{(1,0)y}^{n}\\
		-o_{(0,1)y}^{n}- u_{(1,1)}^{1}\partial^{-1}g_{(1,0)y}^{n}+ u_{(1,1)}^{3}\partial^{-1}e_{(1,0)y}^{n}- u_{(0,1)}^{4}\partial^{-1}j_{(0,0)y}^{n}- u_{(0,1)}^{5}\partial^{-1}a_{(0,0)y}^{n} - u_{(0,1)}^{5}\partial^{-1}c_{(0,0)y}^{n} +u_{(1,1)}^{6}\partial^{-1}d_{(1,0)y}^{n}\\
		p_{(1,1)y}^{n}+ u_{(1,1)}^{2}\partial^{-1}j_{(0,0)y}^{n}+ u_{(1,1)}^{3}\partial^{-1}h_{(0,0)y}^{n}- u_{(0,1)}^{4}\partial^{-1}g_{(1,0)y}^{n} -u_{(0,1)}^{5}\partial^{-1}f_{(1,0)y}^{n} + u_{(1,1)}^{6}\partial^{-1}b_{(0,0)y}^{n}+u_{(1,1)}^{6}\partial^{-1}c_{(0,0)y}^{n}
	\end{array}\right)$%
}
\begin{align*}
	=-LP^{n}-\bar{u}\alpha^{n}(t)x+\bar{P}^{n}_{y}.
\end{align*}
We derive the Hamiltonian structure of (\ref{h2}) via the trace identity
 \begin{align*}
	&\frac{\delta}{\delta u}(2a_{(0,0)}^{n+1}+2b_{(0,0)}^{n+1}+2c_{(0,0)}^{n+1})\\&=(\gamma-n)(q_{(1,1)}^{n},r_{(1,1)}^{n},s_{(1,1)}^{n},-2v_{(0,1)}^{n},-2w_{(0,1)}^{n},2z_{(1,1)}^{n},k_{(1,1)}^{n},l_{(1,1)}^{n},m_{(1,1)}^{n},-2n_{(0,1)}^{n},-2o_{(0,1)}^{n},2p_{(1,1)}^{n})^{t}.
\end{align*}
Setting $n=1$, we obtain $\gamma=0$ by direct computation. Therefore, for $P^{n+1}$ in equation (\ref{h2}), we have:
\begin{align*}
	P^{n+1}=\frac{\delta}{\delta u}(\frac{-2}{n+1}(a_{(0,0)}^{n+2}+b_{(0,0)}^{n+2}+c_{(0,0)}^{n+2})).
\end{align*}
Denoting $H^{n+1}=\frac{-2}{n+1}(a_{(0,0)}^{n+2}+b_{(0,0)}^{n+2}+c_{(0,0)}^{n+2})$, equation (\ref{h2}) can be expressed as:
\begin{align*}
	u_{t_{n}}=JP^{n+1}+u_{x}=J\frac{\delta}{\delta u}H^{n+1}+u_{x}=-JLP^{n}-J\bar{u}\alpha^{n}(t)x+J\bar{P}^{n}_{y}+u_{x}.
\end{align*}
In this chapter, we construct another super-integrable hierarchy on the color Lie algebra $\mathfrak{sp}_{1}(6)$. Although the spatial diemension of its elements differs from that of the integrable hierarchy in Section \ref{sec2}, its main part formallt coincides with the super-integrable hierarchy presented in Section \ref{sec2}.

\appendix
\section{The recurrence operator $L=(l_{i,j})_{12\times 12}$}
\label{app1}
\begin{align*}
	&l_{1,1}=-\frac{1}{2}\partial+u_{(1,1)}^{7}\partial^{-1}u_{(1,1)}^{1}-u^{10}\partial^{-1}u_{(0,1)}^{4}-u_{(0,1)}^{11}\partial^{-1}u_{(0,1)}^{5},\\& l_{1,4}=\frac{1}{2}u_{(1,1)}^{7}\partial^{-1}u_{(0,1)}^{4}+\frac{1}{2}u^{10}\partial^{-1}u_{(1,1)}^{2}+\frac{1}{2}u_{(0,1)}^{11}\partial^{-1}u_{(1,1)}^{6},\\& l_{1,5}=\frac{1}{2}u_{(1,1)}^{7}\partial^{-1}u_{(0,1)}^{5}+\frac{1}{2}u^{10}\partial^{-1}u_{(1,1)}^{6}+\frac{1}{2}u_{(0,1)}^{11}\partial^{-1}u_{(1,1)}^{3},\ l_{1,7}=-u_{(1,1)}^{7}\partial^{-1}u_{(1,1)}^{7},\\& l_{1,8}=-u^{10}\partial^{-1}u^{10},\  l_{1,9}=-u_{(0,1)}^{11}\partial^{-1}u_{(0,1)}^{11},\ l_{1,10}=-\frac{1}{2}u_{(1,1)}^{7}\partial^{-1}u^{10}+\frac{1}{2}u^{10}\partial^{-1}u_{(1,1)}^{7},\\& l_{1,11}=-\frac{1}{2}u_{(1,1)}^{7}\partial^{-1}u_{(0,1)}^{11}+\frac{1}{2}u^{11}\partial^{-1}u_{(1,1)}^{7},\  l_{1,12}=-\frac{1}{2}u^{10}\partial^{-1}u_{(0,1)}^{11}-\frac{1}{2}u_{(0,1)}^{11}\partial^{-1}u^{10},\\&l_{2,2}=-\frac{1}{2}\partial+u_{(1,1)}^{8}\partial^{-1}u_{(1,1)}^{2}-u^{10}\partial^{-1}u_{(0,1)}^{4}+u_{(1,1)}^{12}\partial^{-1}u_{(1,1)}^{6},\\& l_{2,4}=\frac{1}{2}u_{(1,1)}^{8}\partial^{-1}u_{(0,1)}^{4}+\frac{1}{2}u^{10}\partial^{-1}u_{(1,1)}^{1}+\frac{1}{2}u_{(1,1)}^{12}\partial^{-1}u_{(0,1)}^{5},\\& l_{2,6}=\frac{1}{2}u_{(1,1)}^{8}\partial^{-1}u_{(1,1)}^{6}-\frac{1}{2}u^{10}\partial^{-1}u_{(0,1)}^{5}+\frac{1}{2}u_{(1,1)}^{12}\partial^{-1}u_{(1,1)}^{3},\ l_{2,7}=-u^{10}\partial^{-1}u^{10},\\& l_{2,8}=-u_{(1,1)}^{8}\partial^{-1}u_{(1,1)}^{8},\  l_{2,9}=-u_{(1,1)}^{12}\partial^{-1}u_{(1,1)}^{12},\  l_{2,10}=-\frac{1}{2}u_{(1,1)}^{8}\partial^{-1}u^{10}+\frac{1}{2}u^{10}\partial^{-1}u_{(1,1)}^{8},\\& l_{2,11}=\frac{1}{2}u^{10}\partial^{-1}u_{(1,1)}^{12}-\frac{1}{2}u_{(1,1)}^{12}\partial^{-1}u^{10},\ l_{2,12}=-\frac{1}{2}u_{(1,1)}^{8}\partial^{-1}u_{(1,1)}^{12}-\frac{1}{2}u_{(1,1)}^{12}\partial^{-1}u_{(1,1)}^{8},\\& l_{3,3}=-\frac{1}{2}\partial+u_{(1,1)}^{9}\partial^{-1}u_{(1,1)}^{3}-u_{(0,1)}^{11}\partial^{-1}u_{(0,1)}^{5}+u_{(1,1)}^{12}\partial^{-1}u_{(1,1)}^{6},\\& l_{3,5}=\frac{1}{2}u_{(1,1)}^{9}\partial^{-1}u_{(0,1)}^{5}+\frac{1}{2}u_{(0,1)}^{11}\partial^{-1}u_{(1,1)}^{1}+\frac{1}{2}u_{(1,1)}^{12}\partial^{-1}u_{(0,1)}^{4},\\& l_{3,6}=\frac{1}{2}u_{(1,1)}^{9}\partial^{-1}u_{(1,1)}^{6}-\frac{1}{2}u_{(0,1)}^{11}\partial^{-1}u_{(0,1)}^{4}+\frac{1}{2}u_{(1,1)}^{12}\partial^{-1}u_{(1,1)}^{2},\  l_{3,7}=-u_{(0,1)}^{11}\partial^{-1}u_{(0,1)}^{11},\\& l_{3,8}=-u_{(1,1)}^{12}\partial^{-1}u_{(1,1)}^{12},\  l_{3,9}=-u_{(1,1)}^{9}\partial^{-1}u_{(1,1)}^{9},\ l_{3,10}=\frac{1}{2}u_{(0,1)}^{11}\partial^{-1}u_{(1,1)}^{12}-\frac{1}{2}u_{(1,1)}^{12}\partial^{-1}u_{(0,1)}^{11},\\&  l_{3,11}=-\frac{1}{2}u_{(1,1)}^{9}\partial^{-1}u_{(0,1)}^{11}+\frac{1}{2}u_{(0,1)}^{11}\partial^{-1}u_{(1,1)}^{9},\ l_{3,12}=-\frac{1}{2}u_{(1,1)}^{9}\partial^{-1}u_{(1,1)}^{12}-\frac{1}{2}u_{(1,1)}^{12}\partial^{-1}u_{(1,1)}^{9},\\& l_{4,1}=-u_{(1,1)}^{8}\partial^{-1}u_{(0,1)}^{4}-u_{(0,1)}^{10}\partial^{-1}u_{(1,1)}^{1}-u_{(1,1)}^{12}\partial^{-1}u_{(0,1)}^{5},\\& l_{4,2}=-u_{(1,1)}^{7}\partial^{-1}u_{(0,1)}^{4}-u_{(0,1)}^{10}\partial^{-1}u_{(1,1)}^{2}-u_{(0,1)}^{11}\partial^{-1}u_{(1,1)}^{6},\\& l_{4,4}=-\frac{1}{2}\partial+\frac{1}{2}u_{(1,1)}^{8}\partial^{-1}u_{(1,1)}^{2}+u_{(0,1)}^{10}\partial^{-1}u_{(0,1)}^{4}-\frac{1}{2}u_{(0,1)}^{11}\partial^{-1}u_{(0,1)}^{5}+\frac{1}{2}u_{(1,1)}^{12}\partial^{-1}u_{(1,1)}^{6},\\& l_{4,5}=\frac{1}{2}u_{(1,1)}^{8}\partial^{-1}u_{(1,1)}^{3}-\frac{1}{2}u_{(0,1)}^{10}\partial^{-1}u_{(0,1)}^{5}+\frac{1}{2}u_{(1,1)}^{12}\partial^{-1}u_{(1,1)}^{3},\\& l_{4,6}=-\frac{1}{2}u_{(1,1)}^{7}\partial^{-1}u_{(0,1)}^{5}-\frac{1}{2}u_{(0,1)}^{10}\partial^{-1}u_{(1,1)}^{6}-\frac{1}{2}u_{(0,1)}^{11}\partial^{-1}u_{(1,1)}^{3},\\& l_{4,7}=-u_{(1,1)}^{7}\partial^{-1}u_{(0,1)}^{10}+u_{(0,1)}^{10}\partial^{-1}u_{(1,1)}^{7},\ l_{4,8}=-u_{(1,1)}^{8}\partial^{-1}u_{(0,1)}^{10}+u_{(0,1)}^{10}\partial^{-1}u_{(1,1)}^{8},\\& l_{4,10}=\frac{1}{2}u_{(1,1)}^{7}\partial^{-1}u_{(1,1)}^{8}+\frac{1}{2}u_{(1,1)}^{8}\partial^{-1}u_{(1,1)}^{7}+u_{(0,1)}^{10}\partial^{-1}u_{(0,1)}^{10},\\& l_{4,11}=\frac{1}{2}u_{(1,1)}^{7}\partial^{-1}u_{(1,1)}^{12}+\frac{1}{2}u_{(0,1)}^{10}\partial^{-1}u_{(0,1)}^{11}+\frac{1}{2}u_{(0,1)}^{11}\partial^{-1}u_{(0,1)}^{10}+\frac{1}{2}u_{(1,1)}^{12}\partial^{-1}u_{(1,1)}^{7},\\& l_{4,12}=-\frac{1}{2}u_{(1,1)}^{8}\partial^{-1}u_{(0,1)}^{11}+\frac{1}{2}u_{(0,1)}^{10}\partial^{-1}u_{(1,1)}^{12}+\frac{1}{2}u_{(0,1)}^{11}\partial^{-1}u_{(1,1)}^{8}-\frac{1}{2}u_{(1,1)}^{12}\partial^{-1}u_{(0,1)}^{10},\\& l_{5,1}=-u_{(1,1)}^{9}\partial^{-1}u_{(0,1)}^{5}-u_{(0,1)}^{11}\partial^{-1}u_{(1,1)}^{1}-u_{(1,1)}^{12}\partial^{-1}u_{(0,1)}^{4},\\& l_{5,3}=-u_{(1,1)}^{7}\partial^{-1}u_{(0,1)}^{5}-u_{(0,1)}^{10}\partial^{-1}u_{(1,1)}^{6}-u_{(0,1)}^{11}\partial^{-1}u_{(1,1)}^{3},\\& l_{5,4}=\frac{1}{2}u_{(1,1)}^{9}\partial^{-1}u_{(1,1)}^{6}-\frac{1}{2}u_{(0,1)}^{11}\partial^{-1}u_{(0,1)}^{4}+\frac{1}{2}u_{(1,1)}^{12}\partial^{-1}u_{(1,1)}^{2},\\& l_{5,5}=-\frac{1}{2}\partial+\frac{1}{2}u_{(1,1)}^{7}\partial^{-1}u_{(1,1)}^{1}+\frac{1}{2}u_{(1,1)}^{9}\partial^{-1}u_{(1,1)}^{3}-\frac{1}{2}u_{(0,1)}^{10}\partial^{-1}u_{(0,1)}^{4}-u_{(0,1)}^{11}\partial^{-1}u_{(0,1)}^{5}+\frac{1}{2}u_{(1,1)}^{12}\partial^{-1}u_{(1,1)}^{6},\\& l_{5,6}=-\frac{1}{2}u_{(1,1)}^{7}\partial^{-1}u_{(0,1)}^{4}-\frac{1}{2}u_{(0,1)}^{10}\partial^{-1}u_{(1,1)}^{2}-\frac{1}{2}u_{(0,1)}^{11}\partial^{-1}u_{(1,1)}^{6},\\& l_{5,7}=-u_{(1,1)}^{7}\partial^{-1}u_{(0,1)}^{11}+u_{(0,1)}^{11}\partial^{-1}u_{(1,1)}^{7},\ l_{5,8}=u_{(0,1)}^{10}\partial^{-1}u_{(1,1)}^{12}-u_{(1,1)}^{12}\partial^{-1}u_{(0,1)}^{10},\\& l_{5,9}=-u_{(1,1)}^{9}\partial^{-1}u_{(0,1)}^{11}+u_{(0,1)}^{11}\partial^{-1}u_{(1,1)}^{9},\\& l_{5,10}=\frac{1}{2}u_{(1,1)}^{7}\partial^{-1}u_{(1,1)}^{12}+\frac{1}{2}u_{(0,1)}^{10}\partial^{-1}u_{(0,1)}^{11}+\frac{1}{2}u_{(0,1)}^{11}\partial^{-1}u_{(0,1)}^{10}+\frac{1}{2}u_{(1,1)}^{12}\partial^{-1}u_{(1,1)}^{7},\\& l_{5,11}=\frac{1}{2}u_{(1,1)}^{7}\partial^{-1}u_{(1,1)}^{9}+\frac{1}{2}u_{(1,1)}^{9}\partial^{-1}u_{(1,1)}^{7}+u_{(0,1)}^{11}\partial^{-1}u_{(0,1)}^{11},\\& l_{5,12}=-\frac{1}{2}u_{(1,1)}^{9}\partial^{-1}u_{(0,1)}^{10}+\frac{1}{2}u_{(0,1)}^{10}\partial^{-1}u_{(1,1)}^{9}+\frac{1}{2}u_{(0,1)}^{11}\partial^{-1}u_{(1,1)}^{12}-\frac{1}{2}u_{(1,1)}^{12}\partial^{-1}u_{(0,1)}^{11},\\& l_{6,2}=u_{(1,1)}^{9}\partial^{-1}u_{(1,1)}^{6}-u_{(0,1)}^{11}\partial^{-1}u_{(0,1)}^{4}+u_{(1,1)}^{12}\partial^{-1}u_{(1,1)}^{2},\\& l_{6,3}=u_{(1,1)}^{8}\partial^{-1}u_{(1,1)}^{6}-u_{(0,1)}^{10}\partial^{-1}u_{(0,1)}^{5}+u_{(1,1)}^{12}\partial^{-1}u_{(1,1)}^{3},\\& l_{6,4}=\frac{1}{2}u_{(1,1)}^{9}\partial^{-1}u_{(0,1)}^{5}+\frac{1}{2}u_{(0,1)}^{11}\partial^{-1}u_{(1,1)}^{1}+\frac{1}{2}u_{(1,1)}^{12}\partial^{-1}u_{(0,1)}^{4},\\& l_{6,5}=\frac{1}{2}u_{(1,1)}^{8}\partial^{-1}u_{(0,1)}^{4}+\frac{1}{2}u_{(0,1)}^{10}\partial^{-1}u_{(1,1)}^{1}+\frac{1}{2}u_{(1,1)}^{12}\partial^{-1}u_{(0,1)}^{5},\\& l_{6,6}=-\frac{1}{2}\partial+\frac{1}{2}u_{(1,1)}^{8}\partial^{-1}u_{(1,1)}^{2}+\frac{1}{2}u_{(1,1)}^{9}\partial^{-1}u_{(1,1)}^{3}-\frac{1}{2}u_{(0,1)}^{10}\partial^{-1}u_{(0,1)}^{4}-\frac{1}{2}u_{(0,1)}^{11}\partial^{-1}u_{(0,1)}^{5}+u_{(1,1)}^{12}\partial^{-1}u_{(1,1)}^{6},\\& l_{6,7}=-u_{(0,1)}^{10}\partial^{-1}u_{(0,1)}^{11}-u_{(0,1)}^{11}\partial^{-1}u_{(0,1)}^{10},\ l_{6,8}=-u_{(1,1)}^{8}\partial^{-1}u_{(1,1)}^{12}-u_{(1,1)}^{12}\partial^{-1}u_{(1,1)}^{8},\\& l_{6,9}=-u_{(1,1)}^{9}\partial^{-1}u_{(1,1)}^{12}-u_{(1,1)}^{12}\partial^{-1}u_{(1,1)}^{9},\\& l_{6,10}=-\frac{1}{2}u_{(1,1)}^{8}\partial^{-1}u_{(0,1)}^{11}+\frac{1}{2}u_{(0,1)}^{10}\partial^{-1}u_{(1,1)}^{12}+\frac{1}{2}u_{(0,1)}^{11}\partial^{-1}u_{(1,1)}^{8}-\frac{1}{2}u_{(1,1)}^{12}\partial^{-1}u_{(0,1)}^{10},\\& l_{6,11}=-\frac{1}{2}u_{(1,1)}^{9}\partial^{-1}u_{(0,1)}^{10}+\frac{1}{2}u_{(0,1)}^{10}\partial^{-1}u_{(1,1)}^{9}+\frac{1}{2}u_{(0,1)}^{11}\partial^{-1}u_{(1,1)}^{12}-\frac{1}{2}u_{(1,1)}^{12}\partial^{-1}u_{(0,1)}^{11},\\& l_{6,12}=-\frac{1}{2}u_{(1,1)}^{8}\partial^{-1}u_{(1,1)}^{9}-\frac{1}{2}u_{(1,1)}^{9}\partial^{-1}u_{(1,1)}^{8}-u_{(1,1)}^{12}\partial^{-1}u_{(1,1)}^{12},\ l_{7,1}=u_{(1,1)}^{1}\partial^{-1}u_{(1,1)}^{1},\\& l_{7,2}=u_{(0,1)}^{4}\partial^{-1}u_{(0,1)}^{4},\ l_{7,3}=u_{(0,1)}^{5}\partial^{-1}u_{(0,1)}^{5},\  l_{7,4}=\frac{1}{2}u_{(1,1)}^{1}\partial^{-1}u_{(0,1)}^{4}-\frac{1}{2}u_{(0,1)}^{4}\partial^{-1}u_{(1,1)}^{1},\\& l_{7,5}=\frac{1}{2}u_{(1,1)}^{1}\partial^{-1}u_{(0,1)}^{5}-\frac{1}{2}u_{(0,1)}^{5}\partial^{-1}u_{(1,1)}^{1},\  l_{7,6}=\frac{1}{2}u_{(0,1)}^{4}\partial^{-1}u_{(0,1)}^{5}+\frac{1}{2}u_{(0,1)}^{5}\partial^{-1}u_{(0,1)}^{4},\\& l_{7,7}=\frac{1}{2}\partial-u_{(1,1)}^{1}\partial^{-1}u_{(1,1)}^{7}+u_{(0,1)}^{4}\partial^{-1}u_{(0,1)}^{10}+u_{(0,1)}^{5}\partial^{-1}u_{(0,1)}^{11},\\& l_{7,10}=-\frac{1}{2}u_{(1,1)}^{1}\partial^{-1}u_{(0,1)}^{10}-\frac{1}{2}u_{(0,1)}^{4}\partial^{-1}u_{(1,1)}^{8}-\frac{1}{2}u_{(0,1)}^{5}\partial^{-1}u_{(1,1)}^{12},\\& l_{7,11}=-\frac{1}{2}u_{(1,1)}^{1}\partial^{-1}u_{(0,1)}^{11}-\frac{1}{2}u_{(0,1)}^{4}\partial^{-1}u_{(1,1)}^{12}-\frac{1}{2}u_{(0,1)}^{5}\partial^{-1}u_{(1,1)}^{9},\ l_{8,1}=u_{(0,1)}^{4}\partial^{-1}u_{(0,1)}^{4},\\& l_{8,2}=u_{(1,1)}^{2}\partial^{-1}u_{(1,1)}^{2},\ l_{8,3}=u_{(1,1)}^{6}\partial^{-1}u_{(1,1)}^{6},\ l_{8,4}=\frac{1}{2}u_{(1,1)}^{2}\partial^{-1}u_{(0,1)}^{4}-\frac{1}{2}u_{(0,1)}^{4}\partial^{-1}u_{(1,1)}^{2},\\& l_{8,5}=-\frac{1}{2}u_{(0,1)}^{4}\partial^{-1}u_{(1,1)}^{6}+\frac{1}{2}u_{(1,1)}^{6}\partial^{-1}u_{(0,1)}^{4},\ l_{8,6}=\frac{1}{2}u_{(1,1)}^{2}\partial^{-1}u_{(1,1)}^{6}+\frac{1}{2}u_{(1,1)}^{6}\partial^{-1}u_{(1,1)}^{2},\\& l_{8,8}=\frac{1}{2}\partial-u_{(1,1)}^{2}\partial^{-1}u_{(1,1)}^{8}+u_{(0,1)}^{4}\partial^{-1}u_{(0,1)}^{10}-u_{(1,1)}^{6}\partial^{-1}u_{(1,1)}^{12},\\& l_{8,10}=-\frac{1}{2}u_{(1,1)}^{2}\partial^{-1}u_{(0,1)}^{10}-\frac{1}{2}u_{(0,1)}^{4}\partial^{-1}u_{(1,1)}^{7}-\frac{1}{2}u_{(1,1)}^{6}\partial^{-1}u_{(0,1)}^{11},\\& l_{8,12}=-\frac{1}{2}u_{(1,1)}^{2}\partial^{-1}u_{(1,1)}^{12}+\frac{1}{2}u_{(0,1)}^{4}\partial^{-1}u_{(0,1)}^{11}-\frac{1}{2}u_{(1,1)}^{6}\partial^{-1}u_{(1,1)}^{9},\ l_{9,1}=u_{(0,1)}^{5}\partial^{-1}u_{(0,1)}^{5},\\& l_{9,2}=u_{(1,1)}^{6}\partial^{-1}u_{(1,1)}^{6},\ l_{9,3}=u_{(1,1)}^{3}\partial^{-1}u_{(1,1)}^{3},\  l_{9,4}=-\frac{1}{2}u_{(0,1)}^{5}\partial^{-1}u_{(1,1)}^{6}+\frac{1}{2}u_{(1,1)}^{6}\partial^{-1}u_{(0,1)}^{5},\\& l_{9,5}=\frac{1}{2}u_{(1,1)}^{3}\partial^{-1}u_{(0,1)}^{5}-\frac{1}{2}u_{(0,1)}^{5}\partial^{-1}u_{(1,1)}^{3},\ l_{9,6}=\frac{1}{2}u_{(1,1)}^{3}\partial^{-1}u_{(1,1)}^{6}+\frac{1}{2}u_{(1,1)}^{6}\partial^{-1}u_{(1,1)}^{3},\\& l_{9,9}=\frac{1}{2}\partial-u_{(1,1)}^{3}\partial^{-1}u_{(1,1)}^{9}+u_{(0,1)}^{5}\partial^{-1}u_{(0,1)}^{11}-u_{(1,1)}^{6}\partial^{-1}u_{(1,1)}^{12},\\& l_{9,11}=-\frac{1}{2}u_{(1,1)}^{3}\partial^{-1}u_{(0,1)}^{11}-\frac{1}{2}u_{(0,1)}^{5}\partial^{-1}u_{(1,1)}^{7}-\frac{1}{2}u_{(1,1)}^{6}\partial^{-1}u_{(0,1)}^{10},\\& l_{9,12}=-\frac{1}{2}u_{(1,1)}^{3}\partial^{-1}u_{(1,1)}^{12}+\frac{1}{2}u_{(0,1)}^{5}\partial^{-1}u_{(0,1)}^{10}-\frac{1}{2}u_{(1,1)}^{6}\partial^{-1}u_{(1,1)}^{8},\\& l_{10,1}=u_{(1,1)}^{1}\partial^{-1}u_{(0,1)}^{4}-u_{(0,1)}^{4}\partial^{-1}u_{(1,1)}^{1},\ l_{10,2}=u_{(1,1)}^{2}\partial^{-1}u_{(0,1)}^{4}-u_{(0,1)}^{4}\partial^{-1}u_{(1,1)}^{2},\\& l_{10,4}=-\frac{1}{2}u_{(1,1)}^{1}\partial^{-1}u_{(1,1)}^{2}-\frac{1}{2}u_{(1,1)}^{2}\partial^{-1}u_{(1,1)}^{1}-u_{(0,1)}^{4}\partial^{-1}u_{(0,1)}^{4},\\& l_{10,5}=-\frac{1}{2}u_{(1,1)}^{1}\partial^{-1}u_{(1,1)}^{6}-\frac{1}{2}u_{(0,1)}^{4}\partial^{-1}u_{(0,1)}^{5}-\frac{1}{2}u_{(0,1)}^{5}\partial^{-1}u_{(0,1)}^{4}-\frac{1}{2}u_{(1,1)}^{6}\partial^{-1}u_{(1,1)}^{1},\\& l_{10,6}=-\frac{1}{2}u_{(0,1)}^{4}\partial^{-1}u_{(1,1)}^{6}+\frac{1}{2}u_{(1,1)}^{6}\partial^{-1}u_{(0,1)}^{4},\  l_{10,7}=u_{(1,1)}^{2}\partial^{-1}u_{(0,1)}^{10}+u_{(0,1)}^{4}\partial^{-1}u_{(1,1)}^{7}+u_{(1,1)}^{6}\partial^{-1}u_{(0,1)}^{11},\\& l_{10,8}=u_{(1,1)}^{1}\partial^{-1}u_{(0,1)}^{10}+u_{(0,1)}^{4}\partial^{-1}u_{(1,1)}^{8}+u_{(0,1)}^{5}\partial^{-1}u_{(1,1)}^{12},\\& l_{10,10}=\frac{1}{2}\partial-\frac{1}{2}u_{(1,1)}^{1}\partial^{-1}u_{(1,1)}^{7}-\frac{1}{2}u_{(1,1)}^{2}\partial^{-1}u_{(1,1)}^{8}+u_{(0,1)}^{4}\partial^{-1}u_{(0,1)}^{10}+\frac{1}{2}u_{(0,1)}^{5}\partial^{-1}u_{(0,1)}^{11}-\frac{1}{2}u_{(1,1)}^{6}\partial^{-1}u_{(1,1)}^{12},\\& l_{10,11}=-\frac{1}{2}u_{(1,1)}^{2}\partial^{-1}u_{(1,1)}^{12}+\frac{1}{2}u_{(0,1)}^{4}\partial^{-1}u_{(0,1)}^{11}-\frac{1}{2}u_{(1,1)}^{6}\partial^{-1}u_{(1,1)}^{9},\\& l_{10,12}=\frac{1}{2}u_{(1,1)}^{1}\partial^{-1}u_{(0,1)}^{11}+\frac{1}{2}u_{(0,1)}^{4}\partial^{-1}u_{(1,1)}^{12}+\frac{1}{2}u_{(0,1)}^{5}\partial^{-1}u_{(1,1)}^{9},\\& l_{11,1}=u_{(1,1)}^{1}\partial^{-1}u_{(0,1)}^{5}-u_{(0,1)}^{5}\partial^{-1}u_{(1,1)}^{1},\ l_{11,2}=-u_{(0,1)}^{4}\partial^{-1}u_{(1,1)}^{6}+u_{(1,1)}^{6}\partial^{-1}u_{(0,1)}^{4},\\& l_{11,3}=u_{(1,1)}^{3}\partial^{-1}u_{(0,1)}^{5}-u_{(0,1)}^{5}\partial^{-1}u_{(1,1)}^{3},\\& l_{11,4}=-\frac{1}{2}u_{(1,1)}^{1}\partial^{-1}u_{(1,1)}^{6}-\frac{1}{2}u_{(0,1)}^{4}\partial^{-1}u_{(0,1)}^{5}-\frac{1}{2}u_{(0,1)}^{5}\partial^{-1}u_{(0,1)}^{4}-\frac{1}{2}u_{(1,1)}^{6}\partial^{-1}u_{(1,1)}^{1},\\& l_{11,5}=-\frac{1}{2}u_{(1,1)}^{1}\partial^{-1}u_{(1,1)}^{3}-\frac{1}{2}u_{(1,1)}^{3}\partial^{-1}u_{(1,1)}^{1}-u_{(0,1)}^{5}\partial^{-1}u_{(0,1)}^{5},\\& l_{11,6}=\frac{1}{2}u_{(1,1)}^{3}\partial^{-1}u_{(0,1)}^{4}-\frac{1}{2}u_{(0,1)}^{4}\partial^{-1}u_{(1,1)}^{3}-\frac{1}{2}u_{(0,1)}^{5}\partial^{-1}u_{(1,1)}^{6}+\frac{1}{2}u_{(1,1)}^{6}\partial^{-1}u_{(0,1)}^{5},\\& l_{11,7}=u_{(1,1)}^{3}\partial^{-1}u_{(0,1)}^{11}+u_{(0,1)}^{5}\partial^{-1}u_{(1,1)}^{7}+u_{(1,1)}^{6}\partial^{-1}u_{(0,1)}^{10},\\& l_{11,9}=u_{(1,1)}^{1}\partial^{-1}u_{(0,1)}^{11}+u_{(0,1)}^{4}\partial^{-1}u_{(1,1)}^{12}+u_{(0,1)}^{5}\partial^{-1}u_{(1,1)}^{9},\\& l_{11,10}=-\frac{1}{2}u_{(1,1)}^{3}\partial^{-1}u_{(1,1)}^{12}+\frac{1}{2}u_{(0,1)}^{5}\partial^{-1}u_{(0,1)}^{10}-\frac{1}{2}u_{(1,1)}^{6}\partial^{-1}u_{(1,1)}^{8},\\& l_{11,11}=\frac{1}{2}\partial-\frac{1}{2}u_{(1,1)}^{1}\partial^{-1}u_{(1,1)}^{7}-\frac{1}{2}u_{(1,1)}^{3}\partial^{-1}u_{(1,1)}^{9}+\frac{1}{2}u_{(0,1)}^{4}\partial^{-1}u_{(0,1)}^{10}+u_{(0,1)}^{5}\partial^{-1}u_{(0,1)}^{11}-\frac{1}{2}u_{(1,1)}^{6}\partial^{-1}u_{(1,1)}^{12},\\& l_{11,12}=\frac{1}{2}u_{(1,1)}^{1}\partial^{-1}u_{(0,1)}^{10}+\frac{1}{2}u_{(0,1)}^{4}\partial^{-1}u_{(1,1)}^{8}+\frac{1}{2}u_{(0,1)}^{5}\partial^{-1}u_{(1,1)}^{12},\ l_{12,1}=u_{(0,1)}^{4}\partial^{-1}u_{(0,1)}^{5}+u_{(0,1)}^{5}\partial^{-1}u_{(0,1)}^{4},\\& l_{12,2}=u_{(1,1)}^{2}\partial^{-1}u_{(1,1)}^{6}+u_{(1,1)}^{6}\partial^{-1}u_{(1,1)}^{2},\ l_{12,3}=u_{(1,1)}^{3}\partial^{-1}u_{(1,1)}^{6}+u_{(1,1)}^{6}\partial^{-1}u_{(1,1)}^{3},\\& l_{12,4}=\frac{1}{2}u_{(1,1)}^{2}\partial^{-1}u_{(0,1)}^{5}-\frac{1}{2}u_{(0,1)}^{4}\partial^{-1}u_{(1,1)}^{6}-\frac{1}{2}u_{(0,1)}^{5}\partial^{-1}u_{(1,1)}^{2}+\frac{1}{2}u_{(1,1)}^{6}\partial^{-1}u_{(0,1)}^{4},\\& l_{12,5}=\frac{1}{2}u_{(1,1)}^{3}\partial^{-1}u_{(0,1)}^{4}-\frac{1}{2}u_{(0,1)}^{4}\partial^{-1}u_{(1,1)}^{3}-\frac{1}{2}u_{(0,1)}^{5}\partial^{-1}u_{(1,1)}^{6}+\frac{1}{2}u_{(1,1)}^{6}\partial^{-1}u_{(0,1)}^{5},\\& l_{12,6}=\frac{1}{2}u_{(1,1)}^{2}\partial^{-1}u_{(1,1)}^{3}+\frac{1}{2}u_{(1,1)}^{3}\partial^{-1}u_{(1,1)}^{2}+u_{(1,1)}^{6}\partial^{-1}u_{(1,1)}^{6},\\& l_{12,8}=-u_{(1,1)}^{3}\partial^{-1}u_{(1,1)}^{12}+u_{(0,1)}^{5}\partial^{-1}u_{(0,1)}^{10}-u_{(1,1)}^{6}\partial^{-1}u_{(1,1)}^{8},\\& l_{12,9}=-u_{(1,1)}^{2}\partial^{-1}u_{(1,1)}^{12}+u_{(0,1)}^{4}\partial^{-1}u_{(0,1)}^{11}-u_{(1,1)}^{6}\partial^{-1}u_{(1,1)}^{9},\\& l_{12,10}=-\frac{1}{2}u_{(1,1)}^{3}\partial^{-1}u_{(0,1)}^{11}-\frac{1}{2}u_{(0,1)}^{5}\partial^{-1}u_{(1,1)}^{7}-\frac{1}{2}u_{(1,1)}^{6}\partial^{-1}u_{(0,1)}^{10},\\& l_{12,11}=-\frac{1}{2}u_{(1,1)}^{2}\partial^{-1}u_{(0,1)}^{10}-\frac{1}{2}u_{(0,1)}^{4}\partial^{-1}u_{(1,1)}^{7}-\frac{1}{2}u_{(1,1)}^{6}\partial^{-1}u_{(0,1)}^{11},\\& l_{12,12}=\frac{1}{2}\partial-\frac{1}{2}u_{(1,1)}^{2}\partial^{-1}u_{(1,1)}^{8}-\frac{1}{2}u_{(1,1)}^{3}\partial^{-1}u_{(1,1)}^{9}+\frac{1}{2}u_{(0,1)}^{4}\partial^{-1}u_{(0,1)}^{10}+\frac{1}{2}u_{(0,1)}^{5}\partial^{-1}u_{(0,1)}^{11}-u_{(1,1)}^{6}\partial^{-1}u_{(1,1)}^{12}
\end{align*}






\end{document}